\documentclass[11pt, a4paper]{article}
\pdfoutput=1

\usepackage{amsmath,color,multicol}
\usepackage{amsfonts}
\usepackage{amssymb}
\usepackage{graphicx}
\usepackage{mathrsfs}
\usepackage{epsfig}
\usepackage{latexsym}
\usepackage{graphicx}
\usepackage{color}
\usepackage{amsmath,amssymb}
\usepackage{cite}
\usepackage{slashed, cancel}
\usepackage{hyperref}
\usepackage{datetime}
\def\circa#1{\,\raise.3ex\hbox{$#1$\kern-.75em\lower1ex\hbox{$\sim$}}\,}

\newcommand{\beq}{\begin{equation}}
\newcommand{\eeq}{\end{equation}}

\numberwithin{equation}{section}

\font\tenrsfs=rsfs10 at 12pt

\font\sevenrsfs=rsfs7
\font\fiversfs=rsfs5
\newfam\rsfsfam
\textfont\rsfsfam=\tenrsfs
\scriptfont\rsfsfam=\sevenrsfs
\scriptscriptfont\rsfsfam=\fiversfs
\def\mathscr#1{{\fam\rsfsfam\relax#1}}

\hypersetup{colorlinks, citecolor=bluscuro, linkcolor=black, urlcolor=bluscuro}
\definecolor{rossos}{rgb}{0.8,0.2,0.3}
\definecolor{bluscuro}{rgb}{0.15, 0.2, .85}
\definecolor{bluchiaro}{cmyk}{1,.3,0.,0.1}

\makeatother   
\newcommand{\GeV}{{\rm \,GeV}}
\newcommand{\TeV}{{\rm \,TeV}}

\def\de{\textrm{d}}
\def\tr{\textrm{Tr}}

 \def\be   {\begin{equation}}   \def\ee   {\end{equation}}
 \def\ba   {\begin{array}}      \def\ea   {\end{array}}
 \def\bea  {\begin{eqnarray}}   \def\eea  {\end{eqnarray}}
 \def\bean {\begin{eqnarray*}}  \def\eean {\end{eqnarray*}}
 
  \def\noin{\noindent}

\def\hhref#1{\href{http://arxiv.org/abs/#1}{#1}} 

\def\eq#1{eq.~(\ref{#1})}

\begin{document}

%
\begin{flushright} 
CERN-PH-TH/2014-008\hfill
SISSA  01/2014/FISI
\end{flushright}

\vspace{0.5cm}
\begin{center}

{\LARGE \textbf {Benchmarks for Dark Matter Searches at the LHC
%
}}
\\ [1.5cm]
{\large{\textsc{
Andrea De Simone$^{\,a\,}$,
Gian Francesco Giudice$^{\,b,\, c\,}$,
Alessandro Strumia$^{\,d,\, e\,}$
}}}
\\[0.5cm]

\large{\textit{
$^{a}$ SISSA and INFN, Sezione di Trieste, via Bonomea 265, I-34136 Trieste, Italy
\\  \vspace{.5mm}
$^{b}$ CERN, Theory Division, CH-1211 Geneva 23, Switzerland
\\  \vspace{.5mm}
$^{c}$ Solvay Institute, Boulevard du Triomphe, 1050 Bruxelles, Belgium
\\  \vspace{.5mm}
$^d$ Dipartimento di Fisica dell'Universit{\`a} di Pisa and INFN, Italy
\\  \vspace{.5mm}
$^e$ National Institute of Chemical Physics and Biophysics, Tallinn, Estonia
}}
\end{center}

\vspace{0.5cm}

\begin{center}
\textbf{Abstract}
\begin{quote}
We propose some scenarios to pursue dark matter searches at the LHC in a fairly model-independent way.
The first benchmark case is dark matter co-annihilations with coloured particles (gluinos or squarks being special examples).
We determine the masses that lead to the correct thermal relic density including, for the first time,
strong Sommerfeld corrections taking into account colour decomposition.
In the second benchmark case we consider dark matter that couples to SM particles via the $Z$ or the Higgs.
We determine the couplings allowed by present experiments and discuss future prospects.
Finally we present the case of dark matter that freezes out via decays and apply our results to invisible $Z$ and Higgs decays.
\end{quote}
\end{center}

\def\thefootnote{\arabic{footnote}}
\setcounter{footnote}{0}
\pagestyle{empty}

\tableofcontents

\newpage
\pagestyle{plain}

\section{Introduction}

The traditional search for Dark Matter (DM) at the LHC is based on specific theoretical models that are motivated by solving the naturalness problem. Supersymmetry is the prototypical example. The lack of evidence for new physics in the first phase of the LHC, together with the negative results from direct and indirect searches of galactic halo DM, have cast some doubts on the paradigm linking DM to natural electroweak (EW) theories. This has motivated new and more model-independent strategies for DM searches at colliders, and led to a vast literature on the 
subject~\cite{
0403.004,
0503.117,
0808.3384,
0912.4511,
1002.4137,
1003.1912,
1005.1286,
1005.3797,
1008.1591,
1008.1783,
1009.0008,
1011.2310,
1012.2022,
1103.0240,
1103.3289,
1104.1429,
1104.5329,
1107.2048,
1107.2118,
1108.1196,
1108.1800,
1109.3516,
1109.4398,
1111.2359,
1111.2835,
1112.5457,
1201.0506,
1201.3402,
1202.2894,
1203.1662,
1203.3542,
1204.3839,
1206.0640,
1207.1431,
1207.3971,
1208.4361,
1208.4605,
1209.0231,
1210.0195,
1210.0525,
1212.2221,
1212.3352,
1301.1486,
1302.3619,
1303.3348,
1303.6638,
1306.4107,
1307.1129,
1307.2253,
1307.5740,
1307.6277,
1307.8120,
1308.0592,
1308.0612,
1308.2679,
1308.6799,
1310.4491,
1310.6047,
1311.5896,
1311.6169,
1311.7131, 
1312.0009,
1402.1275,
0011335,
0405097,
0605188,
0808.0255,
0811.0393,
0909.0520,
0912.4722,
1005.5651,
1008.1796,
1102.3024,
1106.3097,
1108.0671,
1109.4872,
1110.4405,
1111.4482,
1112.3299,
1201.4814,
1203.2064,
1205.3169,
1309.3561,
1311.1511,
1312.2592, 
1312.5281,
1401.0221,
1402.1173}.
\medskip  

A common approach used to describe the unknown interactions between DM and SM particles is resorting to a set of effective 
operators~\cite{
0808.3384,
0912.4511,
1002.4137,
1005.1286,
1005.3797,
1008.1591,
1008.1783,
1012.2022,
1103.0240,
1104.1429,
1108.1196,
1109.4398,
1201.3402,
1203.1662,
1207.3971,
1208.4361,
1210.0195,
1210.0525,
1212.3352,
1301.1486,
1302.3619,
1303.3348,
1303.6638,
1307.1129,
1307.5740,
1311.7131,
1402.1173}. 
An example is a four-fermion interaction between a spin-1/2 DM particle ($\chi$) and quarks of the kind $(\bar q \gamma_\mu q)(\chi \gamma^\mu \chi)/\Lambda^2$, where  $\Lambda$ is an effective energy scale. At first sight, this approach appears to be fully model-independent, although in practice it has limited 
validity~\cite{
1111.2359,
1112.5457,
1307.2253,
1307.6277,
1308.6799,
1402.1275}. 
At the LHC, the signature is missing energy (from DM) accompanied by a single jet, photon, or $Z$ (required for tagging the event). The signal rate, after the cuts necessary to reduce the SM background, is rather small. This implies that the scales $\Lambda$ of the effective operators probed by the LHC are often smaller than the energy of the partons involved in the collision ($\sqrt{\hat s}$), thus invalidating the use of an effective field theory. As a result, the interpretation of LHC data in terms of effective operators can lead to erroneous conclusions. It can deceptively overestimate the DM signal, because of spurious enhancements proportional to powers of $\sqrt{\hat s}/\Lambda$. Or it can underestimate the actual reach of the LHC search, when the particle that mediates the effective operator is within the kinematical range and gives a much better collider signal than the ``model-independent" DM particle production. Also, the effective-operator approach leads to LHC bounds on the cross sections relevant for direct DM detection that seem very competitive, but are often only illusory. While the effective-operator approximation can be trusted for the low momentum transfers involved in direct detection, an operator with large dimensionality can misleadingly reward LHC for its high energy.  

\medskip  

An alternative approach is to classify possible mediators of the interactions between DM and SM 
particles~\cite{
1003.1912,
1107.2118,
1109.3516,
1111.2359,
1202.2894,
1204.3839,
1207.1431,
1209.0231,
1212.2221,
1307.8120,
1308.0592,
1308.0612,
1308.2679,
1312.5281,
1401.0221}. 
One class of mediators is given by particles exchanged in the $s$-channel of the DM annihilation process (or, inversely, in the DM production process at colliders). These mediators must be electrically neutral and can have spin 0 or 1. The most popular example is a new vector boson $Z^\prime$. A second class of mediators consists of particles exchanged in the $t$-channel. An interesting possibility for the LHC is that the dominant DM annihilation channel is into a quark-antiquark pair and that the $t$-channel mediator is a colour triplet, which is a scalar or vector (if DM has spin 1/2) or a spinor (if DM has spin 0). Direct DM searches give strong constraints on the mediator interactions, but there are still certain windows of mediator mass and couplings that lead to a correct thermal relic abundance and that can be explored by future LHC runs.

\medskip  

The importance of the hunt for DM and our ignorance of its nature entail that the LHC must pursue a diversified, complete, and model-independent program searching for DM. With this paper, we want to contribute to the subject by proposing alternative approaches for strategies that experiments at the LHC can follow in the investigation of DM. 
In section~\ref{sec:2}, we consider a situation in which the DM thermal relic abundance is determined by co-annihilation with a coloured particle. In section~\ref{sec3}, we study the case in which the DM abundance is determined by the coupling with the $Z$ or Higgs boson. In section~\ref{resann} we analyse the case in which the DM abundance is determined by thermal freeze-out of decays, and apply our results to invisible $Z$ and Higgs decays. Finally, section~\ref{sec:conclusions} contains a summary of our results.

\section{DM co-annihilating with a coloured partner}
\label{sec:2}

We consider the possibility that the DM particle, stabilized by a discrete symmetry, is accompanied by a nearby coloured state $\chi'$,
either in the triplet or octet representation of SU(3)$_c$, which can be either a scalar or a fermion. 
These four situations  are summarised in the following table:
\begin{center}
\begin{tabular}{c|cc}
$\chi'$&  Colour triplet & Colour octet \\
\hline
Scalar  & S3 & S8\\
Fermion & F3 & F8\\
\end{tabular}.
\end{center}
Since we  neglect any interaction between the SM and dark sectors, other than strong interactions, there are only two parameters relevant for our analysis:
the DM mass  $M_{\rm DM}$ and the mass splitting $\Delta M$
of $\chi'$ with respect to the DM.

\subsection{DM relic density}

The relic abundance follows from the standard freeze-out mechanism described
by the Boltzmann equation for the total
number density of the dark system
normalised to the entropy
density $Y=n/s$, as a function of $z=M_{\rm DM}/T$
\be
\frac{\de Y}{\de z}=-f(z) (Y^2-Y_{\rm eq}^2)\,,
\label{BE}
\ee
where  $Y_{\rm eq}$ is the thermal equilibrium value of $Y$ and
%
\be
f(z)\equiv \left(1+\frac{1}{3}\frac{\de\ln g_{*S}}{\de\ln T}\right)\frac{s\langle\sigma v\rangle}{z  H}
\approx  \frac{1}{z^2}\sqrt{\frac{\pi g_*}{45}} M_{\rm Pl} M_{\rm DM} \langle\sigma v\rangle \,,
\ee
$g_*$ ($g_{*S}$)
being the number of degrees of freedom that describes the total energy (entropy) of the thermal system.
As well known, by solving the Boltzmann equations one finds that the
observed DM abundance is reproduced for
\be\label{sigmavthermal}
\langle\sigma v\rangle_{\rm cosmo}=(2.3\pm 0.1)\times 10^{-26}\,\textrm{cm}^3\,\textrm{s}^{-1}
\ee
at $T\approx M_{\rm DM}/25$.
In the following we give explicit results for $\langle\sigma v\rangle$ keeping only the
dominant $s$-wave annihilations, while subleading $p$-wave annihilations are included in our
numerical final result.

The dark system  is composed by the DM particle $\chi$, which negligibly annihilates,
and by the coloured partner $\chi'$ which efficiently self-annihilates via QCD interactions.
The two dark particles are kept in thermal equilibrium among themselves by dark interactions.
 The  same interactions will be responsible for the decay 
of $\chi'$ into $\chi$. We assume for simplicity that one $\chi$ particle is produced
in each $\chi'$ decay.
One can describe this system by a single
Boltzmann equation of the form (\ref{BE}) for the quantity $Y=g_\chi Y_\chi + g_{\chi '} Y_{\chi'}$,
and effective cross section 
\be\label{sigmav}
\langle \sigma v\rangle=
\sigma(\chi'\chi'\to \textrm{SM particles})v
\times R^2 ,
\ee
where
\beq \label{eqR}
R=\frac{g_{\chi'} Y_{\chi'}^{\rm eq}}{g_{\chi} Y_{\chi}^{\rm eq}+g_{\chi'} Y_{\chi'}^{\rm eq}} =
\left[ 1+ \frac{g_\chi}{g_{\chi'} }\frac{\exp(\Delta M/T)}{(1+\Delta M/M_{\rm DM})^{3/2}}
\right]^{-1}\ .\eeq

The annihilation channels for the self-conjugate colour octet and for the colour triplet-antitriplet pair are two gluons ($gg$) and SM quarks ($q \bar q$). The number of degrees of freedom is  $g_{\chi'}=\{6,8,12,16\}$ for \{S3, S8, F3, F8\} respectively. In terms of the quadratic Casimir invariant $C(R)$ and the Dynkin index $T(R)$ of a generic irreducible representation $R$ with generators $T^a$,
\be
\delta_{ij} C(R) =(T^aT^a)_{ij}\,,\qquad 
\delta^{ab}T(R)=\tr (T^aT^b) \,,
\ee
we can express the quartic invariants as
\bea
K_1(R)&\equiv& \tr (T^aT^aT^bT^b)=d(R)C^2(R) \cr
K_2(R)&\equiv& \tr (T^aT^bT^aT^b)=K_1(R)-\frac{d(A)C(A)T(R)}{2} \, .
\eea
Here $d(R)$ is the dimensionality of the irreducible representation $R$, and $A$ refers to the adjoint representation. For the fundamental and adjoint of SU($N$), one has
\beq\hbox{
\begin{tabular}{c|ccccc}
$R$ & $d$  & $T$ & $C$ & $K_1$ & $K_2$\\
\hline
fundamental & $N$ & $\frac 12$ & $\frac{N^2-1}{2N}$ & $\frac{(N^2-1)^2}{4N}$ & $-\frac{N^2-1}{4N}$ \\
adjoint & $N^2-1$ & $N$ & $N$ & $N^2(N^2-1)$ & $\frac{N^2(N^2-1)}{2}$ 
\end{tabular}}~~ .
\eeq
Thus, in the case of interest, we find
\beq\hbox{
\begin{tabular}{c|ccccc}
$\chi^\prime$ & $d$  & $T$ & $C$ & $K_1$ & $K_2$\\
\hline
S3, F3 & $3$ & $\frac 12$ & $\frac 43$ & $\frac {16}3$ & $-\frac 23$ \\
S8, F8 & $8$ & $3$ & $3$ & $72$ & $36$ 
\end{tabular}}~~ .
\eeq


In the non-relativistic limit, the annihilation cross sections into gluons is given by\footnote{These formul\ae{} agree with those of~\cite{MDM} taking into account that the parameter $\langle \sigma_A v\rangle$
there defined equals 2 times the conventional $\langle \sigma v\rangle_{\rm cosmo}$ here employed.}
\be
\sigma(\chi'\chi'\to g g )v=\frac{(K_1+K_2)}{16\pi g_{\chi^\prime}d}\frac{g_3^4}{M_{\chi^\prime}^2} \,,
\ee
where $g_3$ is the QCD coupling and $M_{\chi^\prime}$ is the mass of $\chi^\prime$. The annihilation cross section into the six SM quarks (taken to be massless) is
\be
\sigma(\chi'\chi'\to q\bar q)v=\frac{3 T(R)}{\pi  g_{\chi^\prime}d}\frac{g_3^4}{M_{\chi^\prime}^2}\times\left\{
\begin{array}{ll}
1 & \hbox{if $\chi'$ is a fermion}\\
0 & \hbox{if $\chi'$ is a boson}
\end{array}\right.
\,.
\label{ffbar}\eeq
We neglected the electroweak  contributions to the annihilation cross sections
with respect to the dominant QCD effects.
If $\chi'$ is a scalar (S3, S8), its annihilations into
fermions are $p$-wave suppressed. 

Summarising, in the four cases of interest, the total $\chi'$ annihilation cross sections are
\be
\sigma(\chi'\chi'\to gg, q\bar q ) v=\dfrac{g_3^4}{M_{\chi^\prime}^2}\times
\left\{
\begin{array}{ll}
\dfrac{7}{432\pi}& \qquad\textrm{(scalar triplet)}\\[2mm]
\dfrac{27}{256\pi}& \qquad\textrm{(scalar octet)}\\[2mm]
\dfrac{7}{864\pi}+\dfrac{1}{24\pi}& \qquad\textrm{(fermion triplet)}\\[2mm]
\dfrac{27}{512\pi}+\dfrac{9}{128\pi}& \qquad\textrm{(fermion octet)}\\
\end{array}
\right. \,,
\label{crisscross}
\ee
where we have kept separated the contributions from annihilations into gluons and quarks.
The nature of the DM particle enters only in the factor $R$ of eq.~(\ref{eqR}). 
If $\Delta M=0$, then $R= (1+g_\chi /g_{\chi^\prime})^{-1}$ is about equal to one,
as long as the DM number of degrees of freedom is smaller than the one of the coloured $\chi'$ dark state.
For definiteness, we assume that the DM particle is a Majorana fermion,
such as the supersymmetric neutralino, or a complex scalar: $g_\chi=2$ in both cases. From the analytic expression in \eq{crisscross} and the approximate solution in eq.~(\ref{sigmavthermal}), one can easily derive a good first estimate of the relic abundance.

\begin{figure}[t!]
\centering
\includegraphics[width=0.45\textwidth]{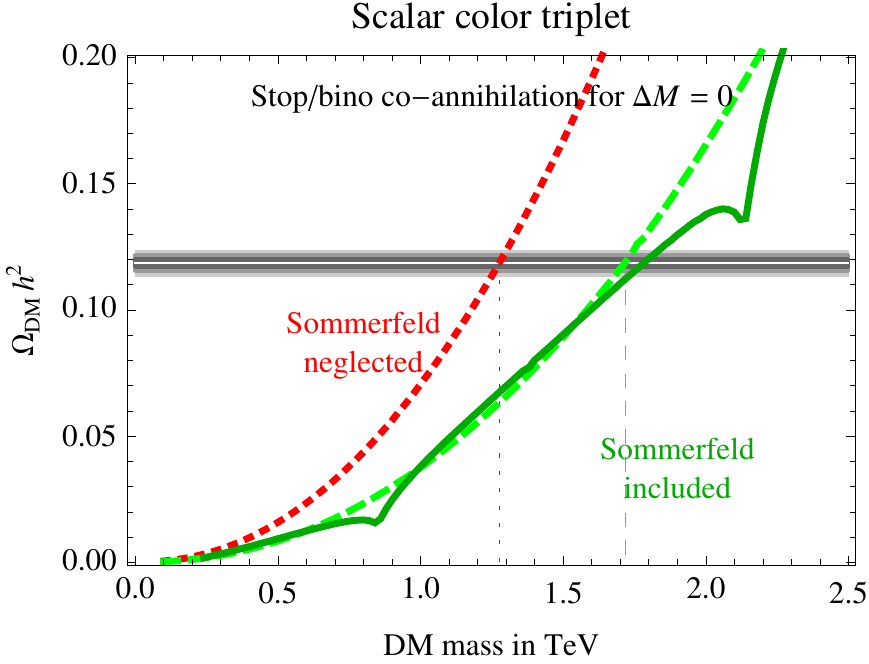}\qquad
\includegraphics[width=0.45\textwidth]{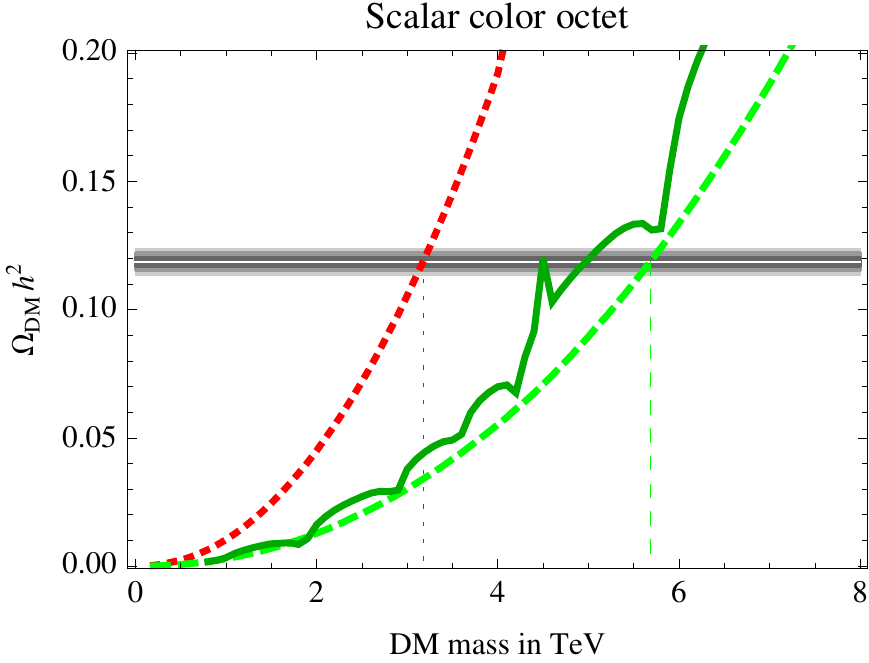}\\[4mm]
\includegraphics[width=0.45\textwidth]{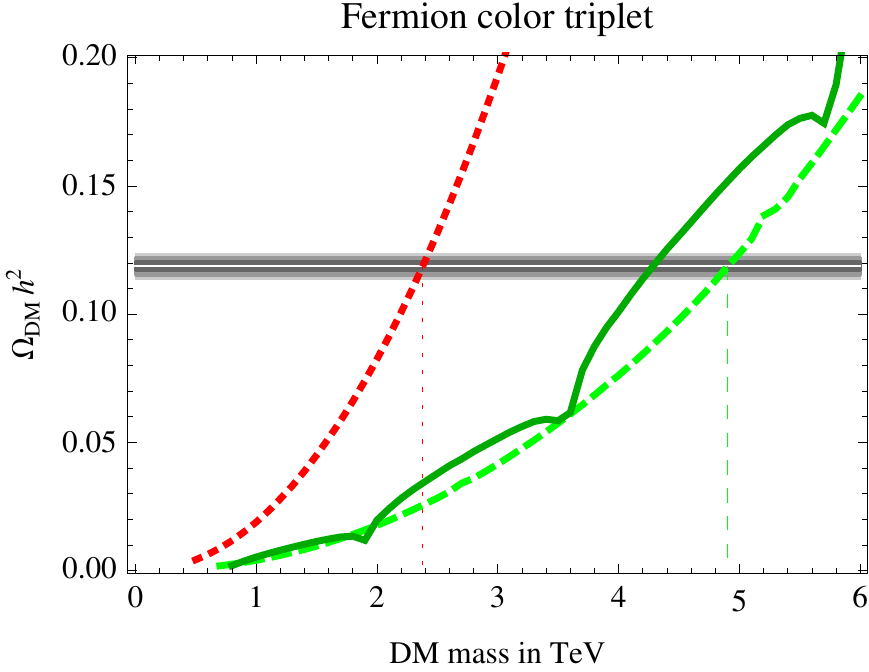}\qquad
\includegraphics[width=0.45\textwidth]{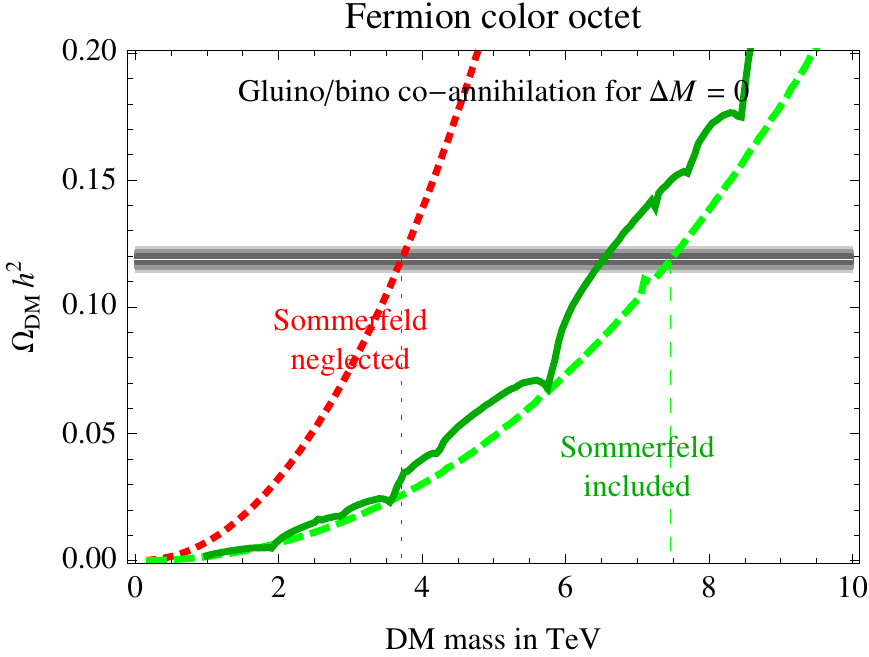}
\caption{\em 
Thermal abundance of DM that co-annihilates with various coloured multiplets
(scalar or fermion, triplet or octet)  in the limit
of mass degeneracy ($\Delta M=0$). Red dashed line: Sommerfeld corrections neglected. Green dashed line: Sommerfeld corrections included analytically. Green solid line: Sommerfeld corrections and gluon thermal mass included numerically.
The horizontal band is the $3\sigma$ experimental range for the DM thermal abundance.} 
\label{fig:cross}
\end{figure}

We have used a numerical solution to the Boltzmann equations, including also $p$-wave annihilations, to obtain the DM thermal relic abundance shown by the red curves in fig.~\ref{fig:cross} (DM abundance as a function of $M_{\rm DM}$ for $\Delta M=0$)
and the red bands in the $(M_{\rm DM},\Delta M)$ plane in fig.~\ref{fig:comb}
(values of
$M_{\rm DM}$ and of $\Delta M$ that correspond to a thermal
DM density equal to the observed cosmological density).

\bigskip

However, the tree level annihilation cross sections discussed so far get substantial Sommerfeld corrections
due to soft-gluon exchanges between the non-relativistic initial states, as we are now going to describe.
After including these corrections, the red curves and bands will shift to the green curves and bands.

\subsection{QCD Sommerfeld  corrections to DM annihilations}
\label{subsec:sommerfeld}

The annihilation of two coloured particles, in the non-relativistic limit,
is strongly affected by  non-perturbative Sommerfeld QCD corrections~\cite{Sommerfeld} (see 
also~\cite{
9806361,
0307216,
0412403,
Strumia:2008cf,
0810.0713,
1005.4678}),
which describe initial state attraction or repulsion due the strong force.
Such corrections have been considered in previous works for gluino annihilation~\cite{9806361} and for stop co-annihilation~\cite{Freitas,Hryczuk};
however we will find a different result and we will not restrict our attention to the  supersymmetric context.

\medskip

We recall that for a single abelian massless vector with potential $V=\alpha/r$, the Sommerfeld correction 
$\sigma_{\rm Sommerfeld} = S \sigma_{\rm perturbative}$ is given by~\cite{Sommerfeld} 
\beq  \label{eq:R0}
S(x) = \frac{-\pi x}{1-e^{\pi x}} \qquad x = \frac{\alpha}{\beta}\, , \eeq
where $\beta$ is the velocity of the incoming particle.
Here $\alpha<0$ describes an attractive potential that leads to an enhancement $S> 1$, and
$\alpha>0$ describes a repulsive potential that leads to $S<1$.
At higher orders, the QCD potential is roughly given by the tree level potential with the strong coupling renormalised at 
the RGE scale  $\bar\mu\approx 1/r$~\cite{VQCD1,VQCD2}
\beq \label{eq:VQCD}
V (r) = C \frac{\alpha_3(\bar\mu)}{r} \bigg[1 + \frac{\alpha_3}{4\pi} \bigg(\frac{11}{7}+14 (\gamma_E + \ln \bar\mu r)\bigg)\bigg]\approx
C \frac{\alpha_3(\bar\mu\approx 1/r)}{r}.\eeq


\noin
The constant $C$ is related to the quadratic Casimir of the two particles involved in the interaction. Using a matrix notation, the QCD potential between two particles (scalar or fermions)
in the representation $R, R'$ of colour  SU(3)$_c$ with generators $T_R^a$ and $T_{R'}^a$ is
\be
V=\frac{\alpha_3}{r}\sum_a T_R^a \otimes T_{R'}^a\,.
\ee
Following ref.~\cite{Strumia:2008cf}, the non-abelian Sommerfeld effect can be reduced to
a combination of abelian-like Sommerfeld corrections, using group theory decompositions.
The non-abelian matrix potential is diagonalised by decomposing the product representation into a sum of irreducible representations $Q$ as
$R\otimes R'=\sum_Q Q$:
\be
V=\frac{\alpha_3}{2r}\left[\sum_Q C_Q {\bf 1}_Q- C_R{\bf 1} - C_{R'} {\bf 1}\right]\,,
\ee
where $C_i$ is the quadratic Casimir of the representation $i$.
The relevant Casimir for our purposes are $C_1=0, C_3=4/3, C_8=3, C_{10}=C_{\overline{10}}=6, C_{27}=8$.

\subsubsection*{Colour triplet}
Annihilations of two colour triplets are decomposed as $3\otimes \overline 3=1\oplus 8$
and the QCD potential becomes
\be
V=\frac{\alpha_3}{r}\times \left\{
\begin{matrix}
-\frac{4}{3} & \qquad\hbox{(1)}\\
+\frac{1}{6}& \qquad\hbox{(8)}\\
\end{matrix}
\right. \,,
\ee
which is attractive  for the singlet two-body state and repulsive for the octet. 
Next, we need to decompose the total annihilation cross-sections computed in the previous sections into partial cross sections relative to the two-body states.

The perturbative $\hbox{triplet}+\overline{\hbox{triplet}} \to g g$ cross section decomposes into the various sub-channels as: 
$2/7$ in the $1$ state,
$5/7$ in the $8$ state (see appendix~\ref{app}).
Thereby, the Sommerfeld-corrected cross section is
\beq \label{eq:S3}
\frac{\sigma(\hbox{triplet}+\overline{\hbox{triplet}}\to gg)_{\rm Sommerfeld}}{  \sigma(\hbox{triplet}+\overline{\hbox{triplet}}\to gg)_{\rm perturbative}}=
 \frac{2}{7} S(-\frac{4\alpha_3}{3\beta})+ \frac{5}{7} S(\frac{\alpha_3}{6\beta}).\eeq
Furthermore, if the state $\chi'$ is a fermion, it also has $s$-wave annihilations into two SM quarks, such that
\beq \frac{\sigma(\hbox{F3}+\overline{\hbox{F3}}\to q\bar q)_{\rm Sommerfeld}}{  \sigma(\hbox{F3}+\overline{\hbox{F3}}\to q\bar q)_{\rm perturbative}}=  S(\frac{\alpha_3}{6\beta}).\eeq
Indeed, an ultra-relativistic $q\bar q$ pair necessarily has spin $S=1$, so that all the cross section
corresponds to the $8$ initial state.

\bigskip

\subsubsection*{Colour octet}
The product of two colour octets decomposes as $8\otimes 8=1_S\oplus 8_A\oplus 8_S\oplus 10_A\oplus \overline{10}_A\oplus 27_S$, where subscripts indicate $A$ntisymmetric and $S$ymmetric combinations.
The QCD potential can be written as
\be
V=\frac{\alpha_3}{r}\times \left\{
\begin{matrix}
-3 & \qquad{(1_S)}\\
-\frac{3}{2}& \qquad{(8_S, 8_A)}\\
\phantom{-}0 & \qquad{(10_A, \overline{10}_A)}\\
+1 & \qquad{(27_S)}\\
\end{matrix}
\right. \,,
\ee
which is attractive for the singlet and octet two-particle states.
The possible two-body states are classified according to 
\beq (C,S,L) = \hbox{(colour, spin, angular momentum)}.\eeq
We are interested only in the dominant $s$-wave annihilations with $L=0$.
For initial scalar octets the anti-symmetric states cannot be in $s$-wave ($L=0$), so
\beq (8,0)\otimes (8,0) = (1_S,0) \oplus (8_S,0)\oplus (27_S,0)\oplus (\hbox{states with angular momentum $L\neq 0$}).\eeq
For initial fermion octets (such as the supersymmetric gluinos) one has
\beq (8,\frac{1}{2})\otimes (8,\frac12) = (1_S,0) \oplus (8_S,0) \oplus (8_A,1) \oplus (10_A,1)\oplus (\overline{10}_A,1)\oplus (27_S,0)\, .\eeq

With the group-theoretical algebra outlined in appendix~\ref{app}, we find that the perturbative $\hbox{octet}+\hbox{octet} \to g g$ cross section decomposes into the various sub-channels as: 
$1/6$ in the $1_S$ state,
$1/3$ in the $8_S$ state and $1/2$ in the $27_S$ state, while the antisymmetric $8_A$, $10_A$ and $\overline{10}_A$ states do not contribute. Since $C$ is conserved by QCD interactions, 
an 
$s$-wave ($L=0$) initial state of two fermions ($C = (-1)^{L+S}$)
can annihilate into $N_V=2$ vectors ($C = (-1)^{N_V}=+1$) only if $S=0$.
Therefore, a scalar and fermion colour octet have the same  Sommerfeld-corrected cross section into gluons:
\beq \label{eq:S8}
\frac{\sigma(\hbox{octet}+\hbox{octet}\to gg)_{\rm Sommerfeld}}{  \sigma(\hbox{octet}+\hbox{octet}\to gg)_{\rm perturbative}}=
 \frac{1}{6} S(-\frac{3\alpha_3}{\beta})+ \frac{1}{3} S(-\frac{3\alpha_3}{2\beta})+ \frac{1}{2} S(\frac{\alpha_3}{\beta}).\eeq
Furthermore, fermion octets  also have $s$-wave annihilations into SM  quarks, which for ultra-relativistic quarks form a
$(8_A,1)$ initial state, so that one simply has
\beq \frac{\sigma(\hbox{F8}+\hbox{F8}\to q\bar q)_{\rm Sommerfeld}}{  \sigma(\hbox{F8}+\hbox{F8}\to q\bar q)_{\rm perturbative}}=S(-\frac{3\alpha_3}{2\beta}).\eeq

\begin{figure}[t!]
\centering
\begin{tabular}{ccc}
\includegraphics[scale=0.8]{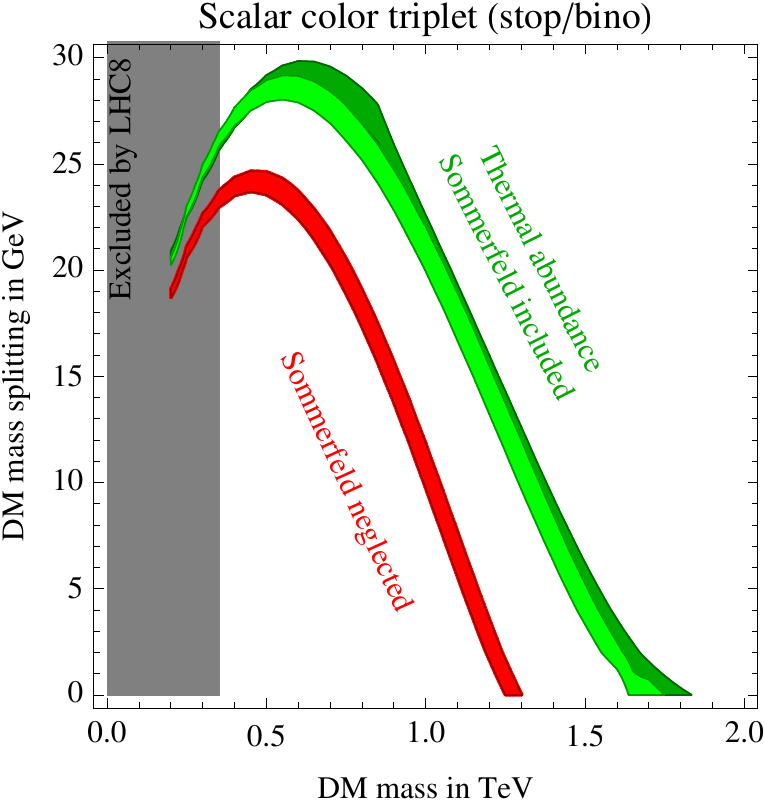}&&
\includegraphics[scale=0.8]{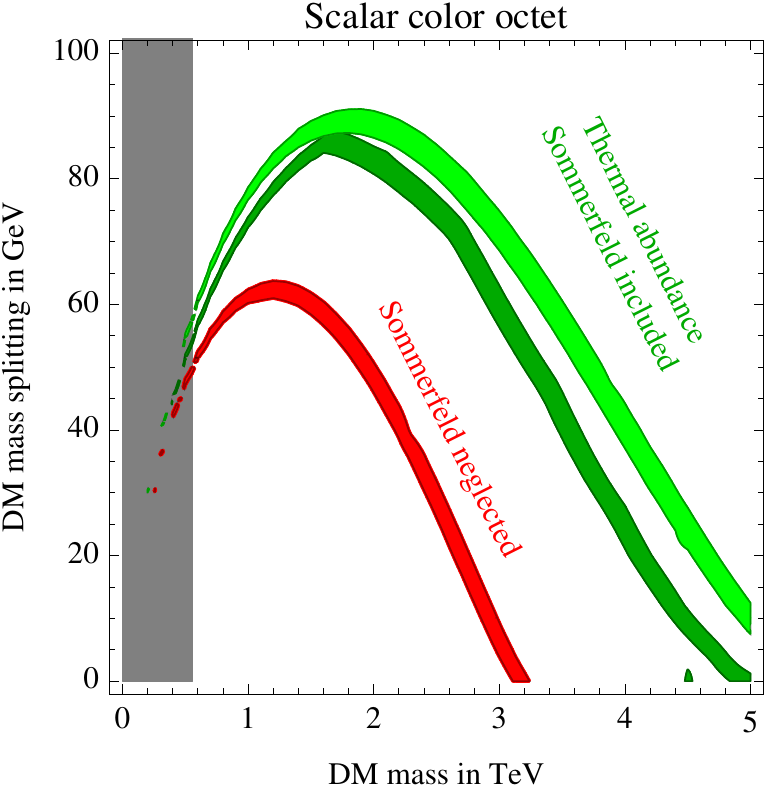}\\
\includegraphics[scale=0.8]{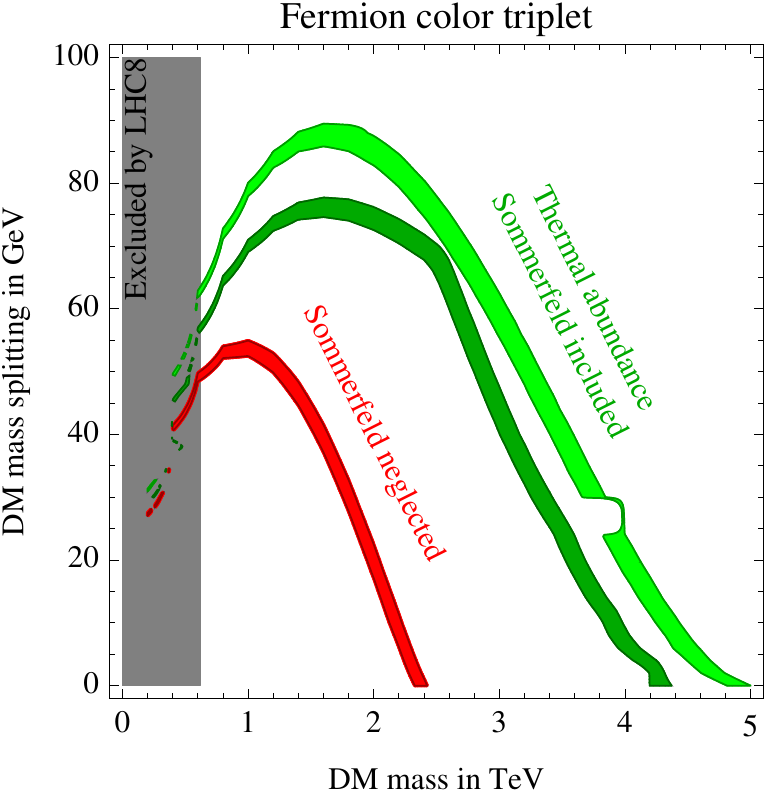}&&
\includegraphics[scale=0.8]{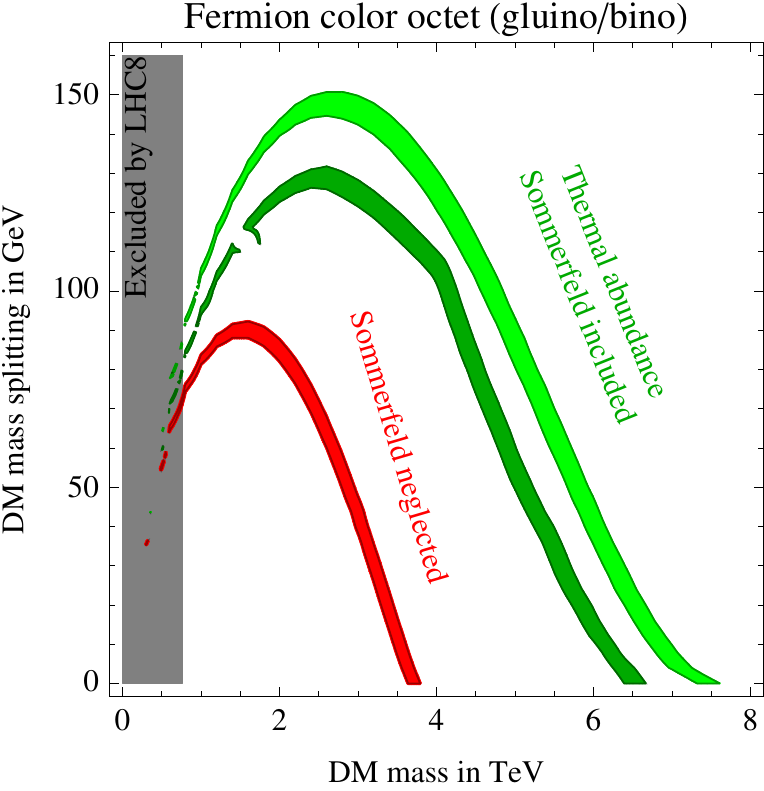}\\
\end{tabular}
\caption{\em
The coloured bands show the region in the $M_{\rm DM}$--$\Delta M$ plane where the correct relic abundance is
achieved for DM co-annihilating with a scalar/fermion colour-triplet/octet partner.
Red: Sommerfeld corrections neglected. Light green: Sommerfeld corrections included analytically. Dark green: Sommerfeld corrections and gluon thermal mass included numerically.
The LHC 90\%CL exclusion is also shown as a vertical grey band.
The DM is assumed to be a Majorana fermion. The case of scalar DM is very similar.
} 
\label{fig:comb}
\end{figure}

\subsection{Results for DM co-annihilations with a coloured partner}
\label{sec:2.3}

By approximating the QCD potential  as proportional to $1/r$
(i.e.\ by renormalising $\alpha_3$ at some fixed relevant scale in eq.~(\ref{eq:VQCD})),
the above equations provide a simple analytical approximation for the Sommerfeld corrections $S$.
In fig.~\ref{fig:comb} we show in light green
the bands in the $(M_{\rm DM}, \Delta M)$ plane where the  DM thermal abundance reproduces the observed
value within $\pm3$ standard deviations.

We also compute the Sommerfeld correction in a more accurate way,
by accounting for two more effects:
\begin{itemize}
\item The full QCD potential taking into account the scale dependence of the QCD coupling constant, as in eq.~(\ref{eq:VQCD});

\item Finite-temperature effects, which provide a zero-momentum thermal mass $m_3 =\sqrt{2} g_3 T$ to the gluon,
transforming the QCD potential into a Yukawa potential.
\end{itemize}
The Coulomb approximation of eq.~(\ref{eq:R0}) no longer holds,
and the Sommerfeld factor $S$ needs to be computed numerically
by  finding the relevant wave-functions with the full potential,
after solving the associated Schr\"odinger equation.

Our full computation leads to the solid green line in fig.~\ref{fig:cross} (relic abundance as a function of $M_{\rm DM}$ for $\Delta M=0$) and to the bands in dark green in fig.~\ref{fig:comb} (relation between $M_{\rm DM}$ and $\Delta M$ that corresponds to DM density).
The irregular peak structure, especially visible in fig.~\ref{fig:cross}, corresponds to the appearance of  
two-body bound states with zero binding energy. Note also that from the results in fig.~\ref{fig:cross} one can easily extract the Sommerfeld-corrected thermal relic abundance of a stable LSP gluino. This is obtained by multiplying the lines shown in the bottom-right panel (fermion colour octet) of fig.~\ref{fig:cross} by the factor $R=8/9$. 

For vanishing mass splitting in the dark sector, $\Delta M=0$,  the Sommerfeld corrections significantly increase the
particle mass needed to reproduce the observed DM abundance:
\beq \begin{array}{c|cccc}
& \hbox{Scalar} & \hbox{Fermion} & \hbox{Scalar} & \hbox{Fermion}\\ 
& \hbox{triplet} & \hbox{triplet} & \hbox{octet} & \hbox{octet}\\ \hline
\hbox{$M_{\rm DM}$ (Sommerfeld neglected)} & 1.3\TeV & 2.4\TeV & 3.2\TeV & 3.7\TeV\\
\hbox{$M_{\rm DM}$ (Sommerfeld included)} & 1.7\TeV & 4-5\TeV & 5-6\TeV & 7-8 \TeV\\
\hbox{LHC lower bound at 90\% CL} & 0.35\TeV & 0.62\TeV & 0.56\TeV & 0.77\TeV\\
\end{array}
\eeq
%
The lower row shows the present bounds from searches at the LHC, computed as follows.
Once produced at the LHC, the coloured $\chi'$ decays into DM (escaping the detector
as missing energy) with the emission
of a jet.
We focus on the case where $\Delta M$ is so low that the radiated jet is too soft to be triggered. 
For larger $\Delta M$, it would be possible to use other search channels with multiple jets and 
missing energy.
The constraints on $\chi'$ production are extracted from the mono-jet DM searches
relying on QCD radiation to trigger the event~\cite{Pierini}.

\smallskip

We simulated with \textsc{MadGraph} \cite{madgraph} 
the tree-level process $pp\to \chi'\chi'$+jet (at $\sqrt{s}=8 \TeV$),
requiring the leading jet to have  $p_T>110 \GeV, |\eta|<2.4$, to reproduce
the analysis in~\cite{monojetATLAS,monojetCMS}, corresponding to data with integrated luminosity of
$\mathcal{L}=19.5$ fb$^{-1}$. 
 Then we used the number of observed and background events
reported for the region with $E_{\rm T}^{\rm miss}>400$ GeV, 
and placed the observed 90\% CL exclusion limit on $M_{\chi'}=M_{\rm DM}+\Delta M$ by requiring
\be
\chi^2=\frac{\left[N_{\rm obs}-N_{\rm bkg}-N_{\chi'}(M_{\rm DM}+\Delta M)\right]^2}
{N_{\chi'}(M_{\rm DM}+\Delta M)+N_{\rm bkg}+\sigma_{\rm bkg}^2}=2.71\,,
\ee
where $\sigma_{\rm bkg}$ is the uncertainty on the background estimation.
We computed the number of signal events $N_{\chi'}$ simply as the integrated luminosity $\mathcal{L}$ times the
signal cross section (with unit efficiency and acceptance). 

Future searches at the LHC will be able to extend the reach for testing DM that co-annihilates with a coloured partner. However, fig.~\ref{fig:comb} shows that the LHC will not be able to probe the entire parameter space allowed by thermal freeze-out. The full exploration of DM co-annihilating with a coloured partner requires higher energies. It is interesting that a future $pp$ collider with $\sqrt{s}\sim 100 \TeV$ will play an important role in the exploration of the mass range favoured by DM thermal abundance~\cite{talk}.





\section{DM annihilating through a SM mediator}\label{sec3}
In this section we consider situations in which the mediator of interactions between
DM and quarks is a SM particle, rather than a speculative
particle from the dark sector.
Given that DM is neutral and has no colour, the candidates for the role of mediator are the $Z$ 
(considered in section~\ref{Z})
and
the Higgs boson (considered in section~\ref{h}).

\subsection{DM coupled to the $Z$}\label{Z}
We start by assuming that the DM particle is coupled to the $Z$ boson.
At low energies, the Lagrangian  interaction of the $Z$ boson  to a current of fermions $f$ and scalars $s$ is
\beq\label{Zcouplings}
{\cal L} =-Z_\mu J^\mu_Z\, , ~~~~ J_\mu^Z = \frac{g_2}{\cos\theta_{\rm W}} \bigg[\sum_f   [\bar{f}\gamma_\mu (g_{V}^f+\gamma_5 g_{A}^f)  f]
+ \sum_s g_s   [s^*( i\partial_\mu s)- (i\partial_\mu s^*) s]\bigg] \, ,
\eeq
where $g_2$ and $\theta_{\rm W}$ are the SU(2)$_L$ gauge coupling and weak angle.
For the SM fermions one has the well-known result
$g_V = \frac{1}{4}-\frac{2}{3} \sin^2\theta_{\rm W}$  and $g_A = -\frac14$ for
up-type quarks and 
$g_V = 
-\frac{1}{4}+\frac{1}{3} \sin^2\theta_{\rm W}$ and $g_A = \frac14$ for down-type quarks.
Since the coupling of each chiral fermion to the $Z$ is proportional to $T_3 - Q \sin^2\theta_{\rm W}=Q\cos^2\theta_{\rm W}-Y$, the coupling of the DM particle, which is neutral ($Q=0$), is proportional to its hypercharge. 
In our effective Lagrangian, we consider $g^{\rm DM}_V$, $g_A^{\rm DM}$ or $g_s^{\rm DM}$
as free parameters that describe the DM couplings.
Small values of the DM couplings to $Z$ can be obtained if the DM is a mixture between a state with $Y=0$ and
a state with $Y\ne 0$, or if DM does not couple directly to $Z$, but only to a $Z'$ boson that mixes with the $Z$.

At energies larger than $M_Z$, we need to complete in a gauge-invariant way the couplings in \eq{Zcouplings}.
This is obtained by observing that, on the Higgs vacuum,
\beq
-\frac{4i\cos\theta_{\rm W}}{g_2\ v^2} \ \left. H^\dagger D_\mu H \right|_{H=\langle H\rangle } =Z_\mu \, ,
\eeq
where $H$ is the full Higgs doublet and $v=246$~GeV. Thus, the simplest gauge invariant completion of the coupling between the $Z$ boson and fermonic or scalar DM  is
\beq\label{ZcouplingsDMPP}
{\cal L} = \frac{4i}{v^2}  (H^\dagger D^\mu H) \bigg [   \bar \psi_{\rm DM}\gamma_\mu (g_V^{\rm DM}+\gamma_5 g_A^{\rm DM})  \psi_{\rm DM}
+ g_s^{\rm DM}   \big(s_{\rm DM}^*( i\partial_\mu s_{\rm DM})- (i\partial_\mu s_{\rm DM}^*) s_{\rm DM}\big)\bigg]\  \, .
\eeq
Indeed, these are the lowest-dimension operators leading to the interactions in \eq{Zcouplings}.

\subsubsection*{Direct detection}
Concerning direct detection, by integrating out the $Z$ at tree level one obtains the effective Lagrangian 
${\cal L}_{\rm eff} = - J_Z^2/2M_Z^2$.
By taking the nucleon matrix element and the non-relativistic limit
we obtain the Lagrangian
\beq {\cal L}_{\rm non~rel} = \sum_N^{n,p} \sum_{i=1}^{12}  c_{i}^N {\cal O}^N_i \, ,
\label{LNR}\eeq
where the first sum runs over $N=\{n,p\}$
and the second sum over the 12 most general DM/nucleon non-relativistic operators~\cite{1008.1591,1203.3542}.
Only a few of them are generated in our case.

\begin{itemize}
\item Scalar DM and fermion DM with vector interactions produces the dominant spin-independent operator ${\cal O}^N_1=1$
with coefficients
\beq c_1^n = 0.27 g^{\rm DM} \frac{M_{\rm DM} m_N}{M_Z^2},\qquad
c_1^p =-0.03 g^{\rm DM} \frac{M_{\rm DM} m_N}{M_Z^2} \, ,
\eeq
where $g$ is either $g_V^{\rm DM}$ or $g_s^{\rm DM}$.  All other operators give negligible corrections.

\medskip

\item
Fermion DM with axial interactions produces, as main effect,  the dominant spin-dependent operator
${\cal O}^N_4= \vec S_{\rm DM}\times\vec s_N$:
\beq c_4^n =c_4^p = 0.38 g_A^{\rm DM} \frac{M_{\rm DM} m_N}{M_Z^2}.\eeq
It also produces ${\cal O}^N_8 =\vec S_{\rm DM}\cdot\vec v_\perp$, which is spin-independent but suppressed by the DM transverse velocity $\vec v_\perp$.  In view of the coefficients
\beq c_8^n = 0.54 g_A^{\rm DM} \frac{M_{\rm DM} m_N}{M_Z^2},\qquad
c_8^p =0.06 g_A^{\rm DM} \frac{M_{\rm DM} m_N}{M_Z^2}\eeq
such operator is somehow less relevant than ${\cal O}^N_4$, but not irrelevant.\footnote{Fermion DM with axial interactions also produces other operators,
${\cal O}^N_7 =\vec s_N\cdot\vec v_\perp$ and
${\cal O}^N_9=i \vec S_{\rm DM}\cdot (\vec{s}_N\times\vec q)$,
which have a negligible effect because spin-dependent and suppressed.}
\end{itemize}

%
%
%
%
%
%
%
%
%
%
%
We derive the bound from all direct detection experiments by employing the public code of ref.~\cite{Cirelli}.
The bounds are dominated by the LUX experiment~\cite{lux}. The effect of
loop corrections that transform spin-dependent interactions into spin-independent cross section is irrelevant, since it can affect our bound on $g_A^{\rm DM}$ only
if DM is lighter than a few GeV~\cite{Haisch}.

\subsubsection*{Thermal abundance}
We compute the relic abundance using the interaction between DM and SM particles given in \eq{ZcouplingsDMPP}.
This interaction contributes to DM annihilation via $s$-channel $Z$ exchange and also to direct annihilation into a pair of Higgs and/or gauge bosons. We perform a full calculation of the relic abundance, including all annihilation channels. The approximation of retaining only the dimension-6 interaction in \eq{ZcouplingsDMPP} is valid as long as the effective energy scale ($v/\sqrt{g_{V,A,s}^{\rm DM}}$) is much larger than the DM mass. This implies $g_{V,A,s}^{\rm DM}\ll 0.24\ (500\GeV/M_{\rm DM})^2$, which is valid in the region of interest. However, if new physics is not far from $M_{\rm DM}$, new interactions and new annihilation channels open up, presumably reducing the thermal relic abundance. These effects are completely model-dependent.


The computation of the thermal relic DM abundance becomes model-independent in the kinematic region
$M_{\rm DM}\approx M_Z/2$, since the annihilation cross section is dominated by 
the $Z$-resonance. We postpone the discussion of this interesting case to section~\ref{resann}, 
where we will show that the DM abundance can be simply computed
in terms of the $Z$ decay width rather than in terms of DM annihilations.



\begin{figure}[t!]
\centering
\includegraphics[height=6cm, width=5cm]{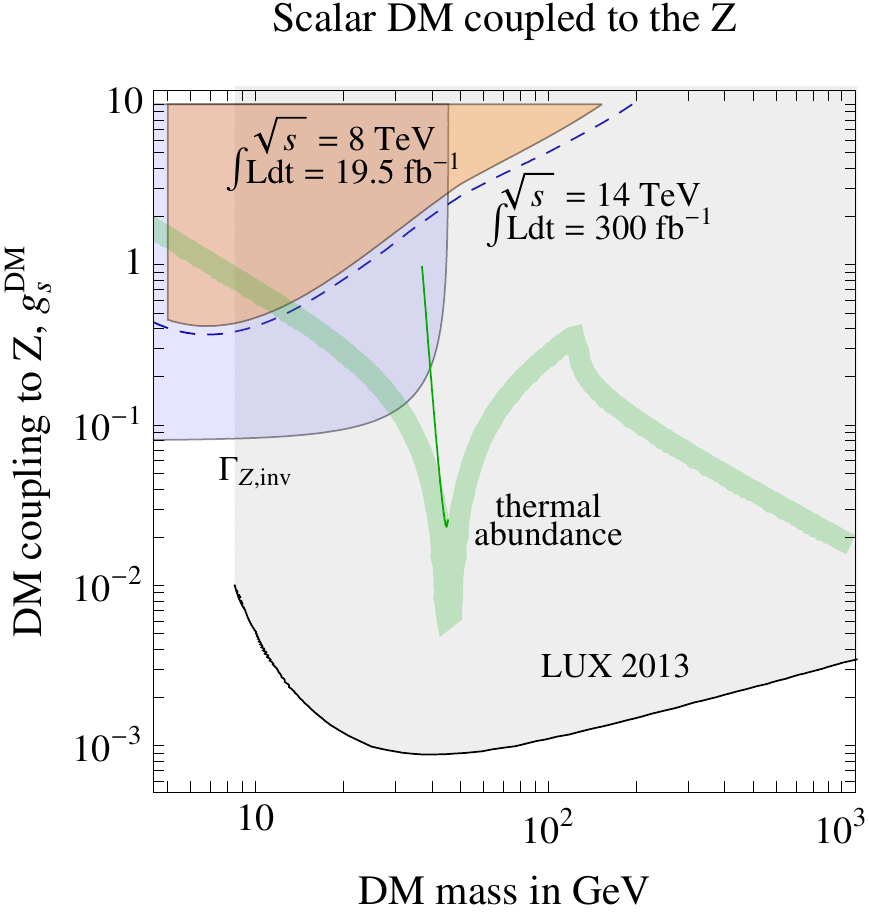}
\hspace{0.3cm}
\includegraphics[height=6cm, width=5cm]{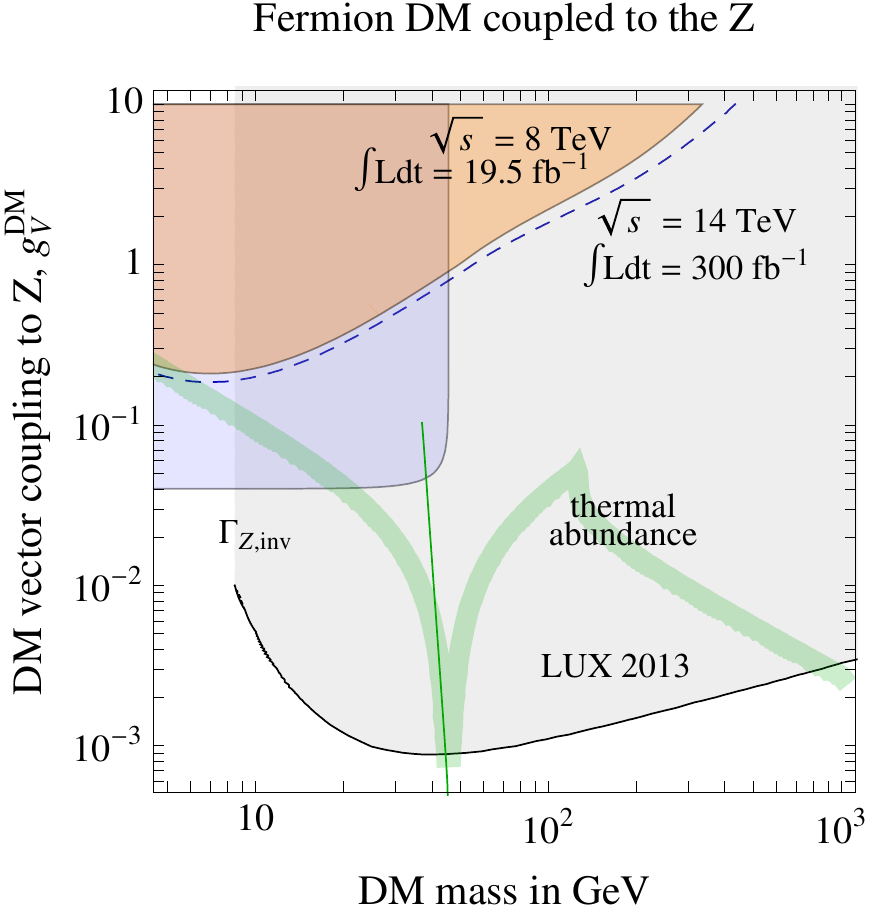}
\hspace{0.3cm}
\includegraphics[height=6cm, width=5cm]{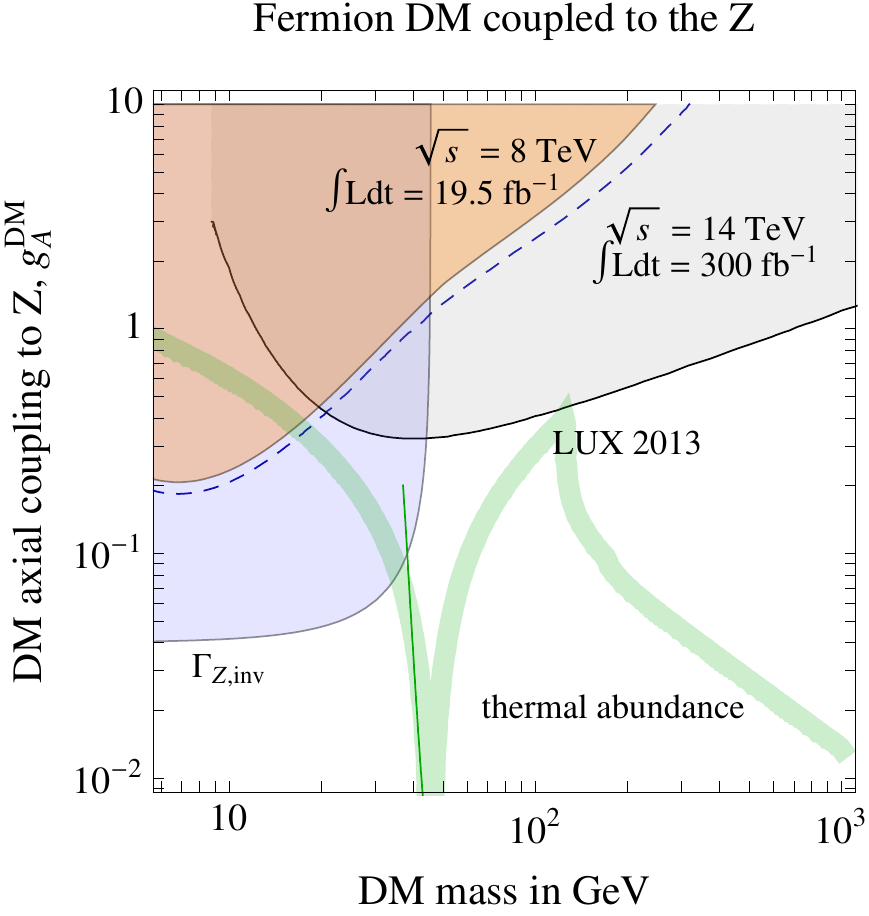}
\caption{\em
\textbf{DM coupled to the $Z$.}
Regions of DM mass $M_{\rm DM}$ and $Z$ couplings ($g_s^{\rm DM}$, $g_V^{\rm DM}$, $g_A^{\rm DM}$):
the orange region is excluded at $90\%$ CL by ATLAS mono-jet searches at LHC8, with forecast for LHC14 (dashed blue line);
the grey region is excluded at $90\%$ CL by LUX 2013 direct searches; the blue region is excluded by the
$Z$-invisible width constraint $\Gamma_{Z, \rm inv}< 2$ {\rm MeV}.
The green solid curve corresponds to a thermal relic abundance via $Z$-coupling annihilation equal to the observed DM density
(the thick curve is the off-shell estimation; the thin curve is the on-shell computation).} 
\label{fig:exclusionsZ}
\end{figure}

\subsubsection*{Results}
In fig.~\ref{fig:exclusionsZ} we compare the LHC sensitivity with the current bounds. 
In the plane (DM mass, DM coupling to $Z$) we show:
\begin{enumerate}
\item The bounds from direct detection, dominated by the LUX experiments (regions shaded in grey).
The bounds on $g_V^{\rm DM}$ and $g_s^{\rm DM}$ are quite strong (around $10^{-3}$ for DM mass around 100~GeV), while $g_A^{\rm DM}$, which leads to spin-dependent interactions, is less constrained (typically $g_A^{\rm DM} \circa{<} 0.3$ for $M_{\rm DM}\approx 100\GeV$). 
We see that direct detection experiments  severely constrain the vector coupling $g_V^{\rm DM}$ and the scalar coupling $g^{\rm DM}_s$, and are presently probing the region $g_A^{\rm DM}\sim 1$.

\item The LEP bounds from the invisible $Z$ width, $\Gamma_{Z, \rm inv}< 2$ MeV.
This bound, shown in light blue, implies $g^{\rm DM}_{V,A}\circa{<} 0.04$, $g^{\rm DM}_{s}\circa{<} 0.08$ if $M_{\rm DM}< M_Z/2$.

\item The present bound from LHC  mono-jet searches, extracted with the procedure described in section~\ref{sec:2.3}.
We see that such bounds can never be competitive with the combined limits from LUX and LEP.

\item Our estimate on the future sensitivity of LHC at $\sqrt{s}=14\TeV$ with an integrated luminosity of 300 fb$^{-1}$. By simulating the sample and rescaling the corresponding statistical error with
the square-root of the number of events we find that only a modest improvement is possible. New strategies for reducing the systematic error and improving background rejection are necessary for the LHC to give competitive results.

\item The curve that corresponds to a thermal DM density equal to the cosmological density (green curve). We observe that a thermal abundance from pure $Z$ coupling is ruled out for scalar DM, while some regions are still allowed for fermion DM, most notably for axial couplings and in the window around the near-resonant region (that will be discussed in section~\ref{resann}). However, we stress that the relic abundance, computed here using the effective interaction in \eq{ZcouplingsDMPP}, is very sensitive to new-physics effects, especially in the high-mass region. In particular, the decrease of the green line with the DM mass is only a consequence of the non-renormalisable contact interactions. New particles and new interactions can completely modify the behaviour of the thermal-abundance constraint. Hence, the green curve in fig.~\ref{fig:exclusionsZ} is only meant to be indicative of the effective-theory regime. 
\end{enumerate}




\subsection{DM coupled to the Higgs}\label{h}
The case of DM that couples to the SM sector only though interactions with the Higgs boson has been discussed extensively in the literature~\cite{
0011335,
0405097,
0605188,
0808.0255,
0811.0393,
0909.0520,
0912.4722,
1005.5651,
1008.1796,
1102.3024,
1106.3097,
1108.0671,
1109.4872,
1110.4405,
1111.4482,
1112.3299,
1201.4814,
1203.2064,
1205.3169,
1309.3561,
1311.1511,
1312.2592}.
Here we assume that DM is either
a real scalar ($s_{\rm DM}$) or a Majorana fermion ($\psi_{\rm DM})$ coupled to the physical Higgs field $h$ at low energies as
\beq 
{\cal L} =-h J_h\, , ~~~~
 J_h =\frac{1}{\sqrt{2}}
 \bigg[ \sum_f  y_f \bar f f  + 
\bar\psi_{\rm DM} (y_{\rm DM}  + i y^P_{\rm DM} \gamma_5)\psi_{\rm DM} +\frac{ \lambda_{\rm DM} v}{2}s^2_{\rm DM}
\bigg].
\label{eq:DMh}
\eeq 
The SM fermions $f$ have the usual Yukawa couplings $y_f$ and
we parameterise the DM couplings to the Higgs as $\lambda_{\rm DM}$, $y_{\rm DM}$, $y_{\rm DM}^P$.

We can complete the effective interaction in \eq{eq:DMh} in a straightforward way, since $H^\dagger H/v = \sqrt{2}h +\dots$. Hence, the simplest recipe to express the DM coupling to Higgs boson in terms of gauge-invariant quantities is
\beq
{\cal L}=-H^\dagger H
 \bigg[ 
\bar\psi_{\rm DM} \frac{(y_{\rm DM}  + i y^P_{\rm DM} \gamma_5)}{2 v}\psi_{\rm DM} +\frac{ \lambda_{\rm DM}}{4}s^2_{\rm DM}
\bigg] \, .
\label{HcoupGI}
\eeq
Note that the coupling of scalar DM to the Higgs doublet can be expressed in terms of a renormalisable interaction, while the coupling of fermonic DM involves a dimension-5 operator.



%
%
%

\subsubsection*{Direct detection}
By integrating out the Higgs boson, one obtains the effective Lagrangian 
${\cal L}_{\rm eff} =  J_h^2/2M_h^2$ that describes direct detection.
Employing again the non-relativistic nucleon Lagrangian of eq.~(\ref{LNR}) we find:
\begin{itemize}
\item  The $\lambda_{\rm DM}$ coupling of scalar DM generates the dominant  spin-independent effective non-relativistic operator
${\cal O}^N_1=1$ with coefficients
\beq c_1^n \approx c_1^p = -0.45\lambda_{\rm DM}  \frac{m_Nv}{M_h^2} .\eeq

\item  The $y_{\rm DM}$ coupling of fermion DM also generates  ${\cal O}^N_1$ with
\beq c_1^n \approx c_1^p = -1.8 y_{\rm DM} \frac{m_NM_{\rm DM}}{M_h^2}\, .
\eeq

\item The pseudo-scalar coupling $y_{\rm DM}^P$ only produces the operator ${\cal O}_{11}^N = i \vec S_{\rm DM}\cdot\vec q$,
which is spin-dependent and suppressed by the transferred momentum $\vec q$:
\beq c_{10}^n\approx  c_{10}^p \approx 0.26  \frac{y_{\rm DM}^Pm_N}{M_h^2}  .\eeq
As a consequence, there are no limits on perturbative values of $y_{\rm DM}^P$.

\end{itemize}

\subsubsection*{Thermal abundance}
The relic abundance is computed using the interaction in \eq{HcoupGI}, which contributes to DM annihilation through $s$-channel Higgs exchange and through processes with two Higgs or longitudinal gauge bosons in the final state. We include these annihilation channels in our computation. 
In the case of fermionic DM, the approximation of keeping only the dimension-5 operator in \eq{HcoupGI} is justified as long as $y_{\rm DM}\ll 0.5\ (500\GeV /M_{\rm DM} )$.


\begin{figure}[t!]
\centering
\includegraphics[height=6cm, width=5cm]{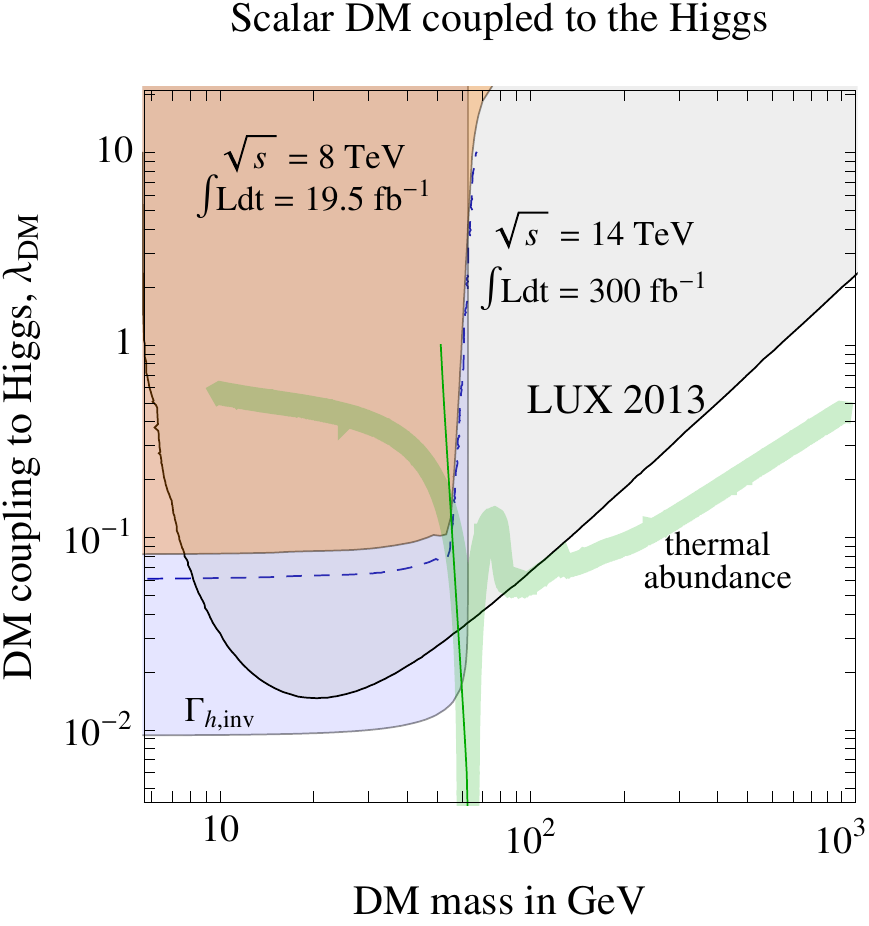}
\hspace{0.3cm}
\includegraphics[height=6cm, width=5cm]{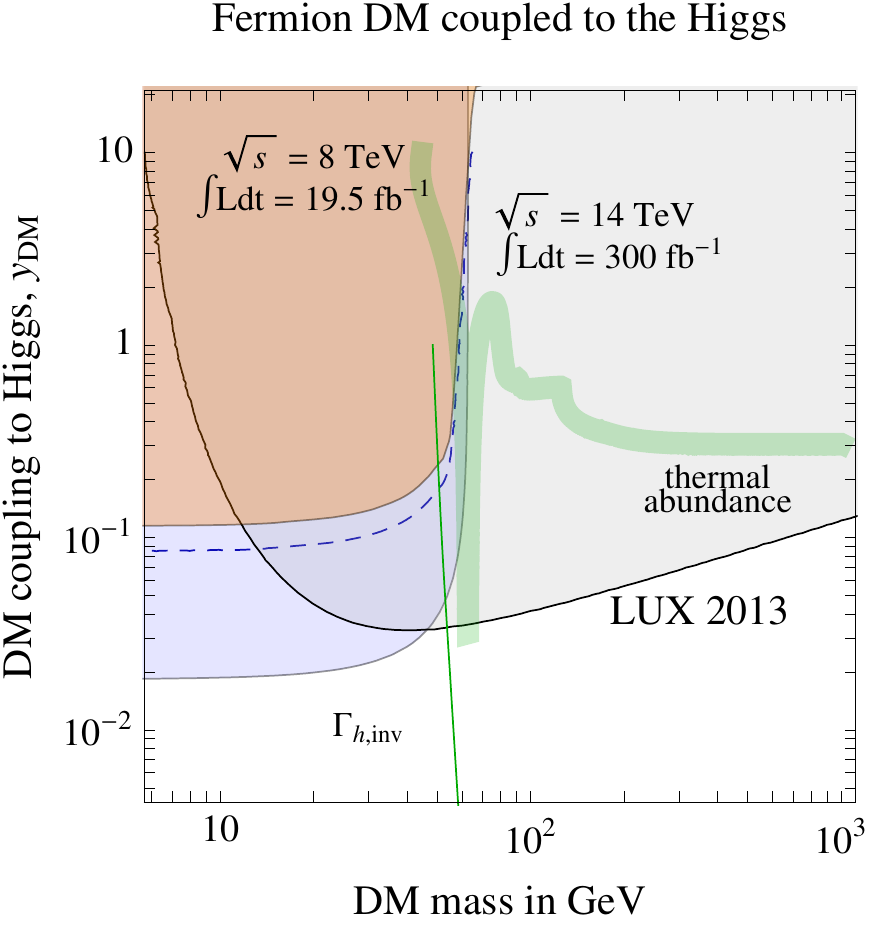}
\hspace{0.3cm}
\includegraphics[height=6cm, width=5cm]{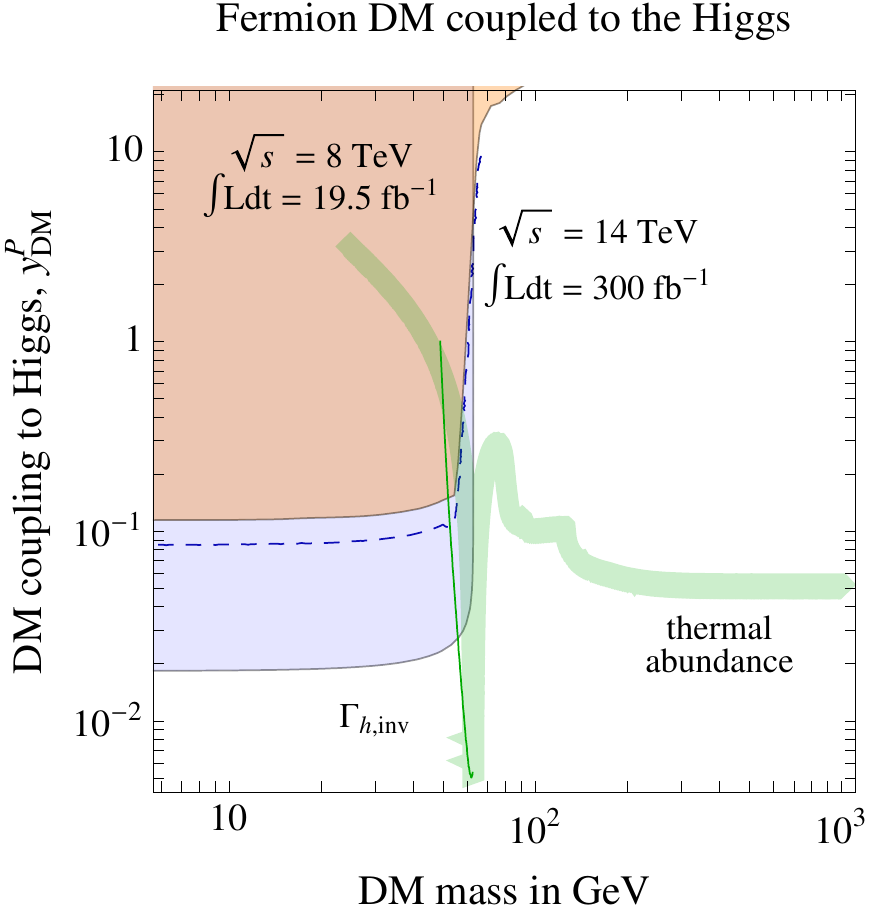}
\caption{\em
\textbf{DM coupled to the Higgs.}
Regions of DM mass $M_{\rm DM}$ and Higgs couplings ($\lambda_{\rm DM}$, $y_{\rm DM}$, $y_{\rm DM}^P$):
the orange region is excluded at $90\%$ CL by ATLAS mono-jet searches at LHC8, with forecast for LHC14 (dashed blue line);
the grey region is excluded at $90\%$ CL by LUX 2013 direct searches; the blue region is excluded by the 
Higgs invisible width constraint $\Gamma_{h, \rm inv}/\Gamma_{h}< 20\%$.
The green solid curve corresponds to a thermal relic abundance via Higgs-coupling annihilation equal to the observed DM density
(the thick curve is the off-shell estimation; the thin curve is the on-shell computation).} 
\label{fig:exclusionsHiggs}
\end{figure}

\subsubsection*{Results}
In fig.~\ref{fig:exclusionsHiggs} we compare the LHC sensitivity with current bounds, 
in the plane (DM mass, DM coupling to $h$), finding the following results. 
\begin{enumerate}
\item 
 The bounds from direct detection are dominated by the LUX experiments (regions shaded in grey).
We see that direct detection experiments are severely constraining the scalar couplings $\lambda_{\rm DM}, y_{\rm DM}$, while the pseudo-scalar interaction is completely out of reach
at the moment.

\item If $M_{\rm DM}<M_h/2$, the main constraint is due to the Higgs invisible width, $\Gamma_{h, \rm inv}/\Gamma_h\lesssim 20\%$,
which gives $\lambda_{\rm DM}, y_{\rm DM}, y_{\rm DM}^P\circa{<}10^{-2}$, 
taking $\Gamma_{h}=4.2$ MeV for $M_h=125.6$ GeV.

\item 
In the opposite regime, $M_{\rm DM}>M_h/2$, one can consider different Higgs production mechanisms
at the LHC:  gluon fusion accompanied by mono-jet, VBF, Higgs-strahlung from $W/Z$.
We considered the first case (gluon fusion) and assumed $M_h=125.6$ GeV. 
However, the parameter space region accessible by LHC mono-jet searches
is either already excluded by direct detection ($\lambda_{\rm DM}, y_{\rm DM}$) or involves unreasonably large couplings ($y_{\rm DM}^P$).

\item As for the case of DM coupling to the $Z$,  the present bound from LHC  mono-jet searches, extracted with the procedure described in section~\ref{sec:2.3}, are not competitive
 with the combined limits from LUX and Higgs invisible width, not even projecting the sensitivity of LHC14 with 300 
 fb$^{-1}$. 

\item 
The case of a DM coupling to the Higgs responsible for the correct relic abundance is ruled out for fermionic DM (but allowed for pseudoscalar coupling when $M_{\rm DM}>M_h/2$). For scalar DM, this possibility is still viable for $M_{\rm DM}\circa{>} 100$~GeV. A small mass window around the resonant Higgs exchange is allowed, and this case will be discussed in section~\ref{resann}.
However, we recall again that the thermal abundance lines in fig.~\ref{fig:exclusionsHiggs} 
bear a dependence on the completion of the theory and our calculation is based on an effective-theory regime with couplings defined by  \eq{HcoupGI}. In particular, for fermonic DM, the green line in fig.~\ref{fig:exclusionsHiggs} is approximately independent of the DM mass in the high-mass region; this result is characteristic of dimension-5 interactions. New particles and new interactions can easily reduce the cosmological abundance of the DM particle coupled to the Higgs.  

\end{enumerate}
%

\section{DM freeze-out via decays}\label{resann}
A special case occurs when the DM annihilation cross section relevant for the thermal relic abundance is resonantly enhanced by the mediator exchange in the $s$-channel. 
This applies when the DM mass is about $M_Z/2=45.6$~GeV or $M_h/2=63$~GeV, but our considerations apply to the case of a generic mediator $M$ (such as extra Higgses present in supersymmetric models or $Z'$ gauge bosons).
We will consider a mediator $M$ with $g_M$ degrees of freedom,
with mass $M_M$ slightly larger than $2M_{\rm DM}$, 
with branching ratio $\hbox{BR}_{\rm DM}$ into a pair of DM particles and $1-\hbox{BR}_{\rm DM}$ into light SM particles.
The DM particle has $g_{\rm DM}$ degrees of freedom.
If the DM mass differs from half of the mediator mass by less than its width, then the cross section 
becomes  fairly model-independent and is approximated
by the Breit-Wigner formula. The thermal average $\gamma_A$ of a resonant annihilation cross section  can be simplified as follows.

\begin{figure}[t!]
\centering
\includegraphics[width=0.45\textwidth]{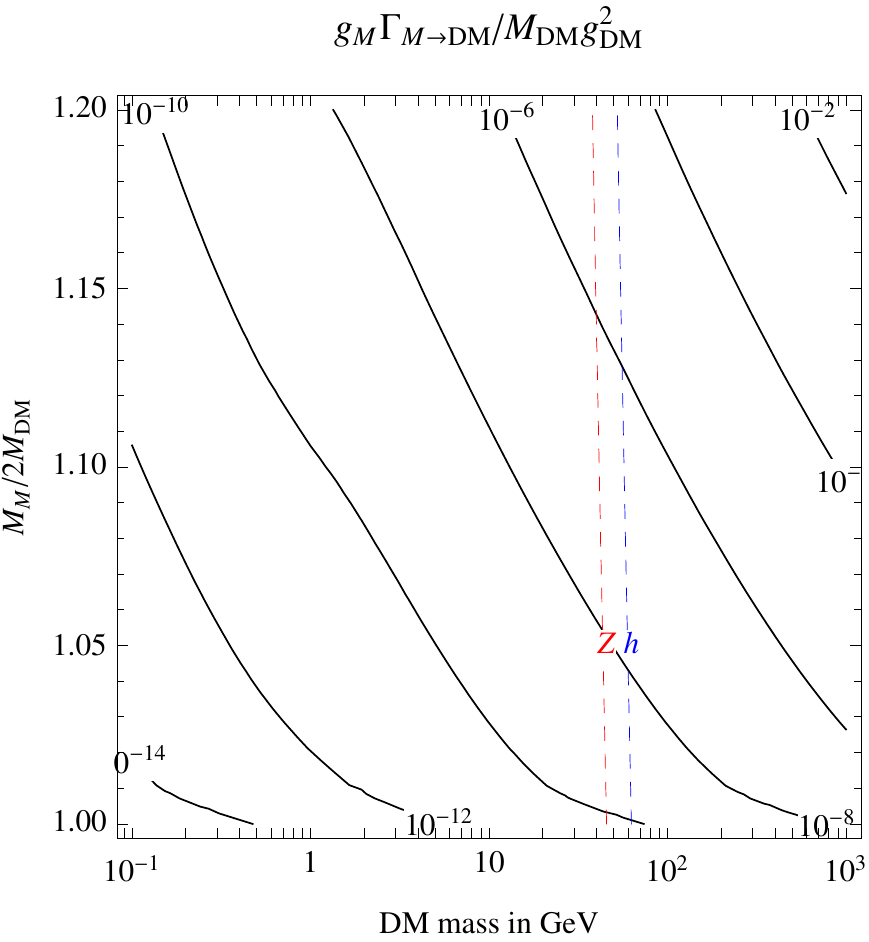}\qquad
\includegraphics[width=0.45\textwidth]{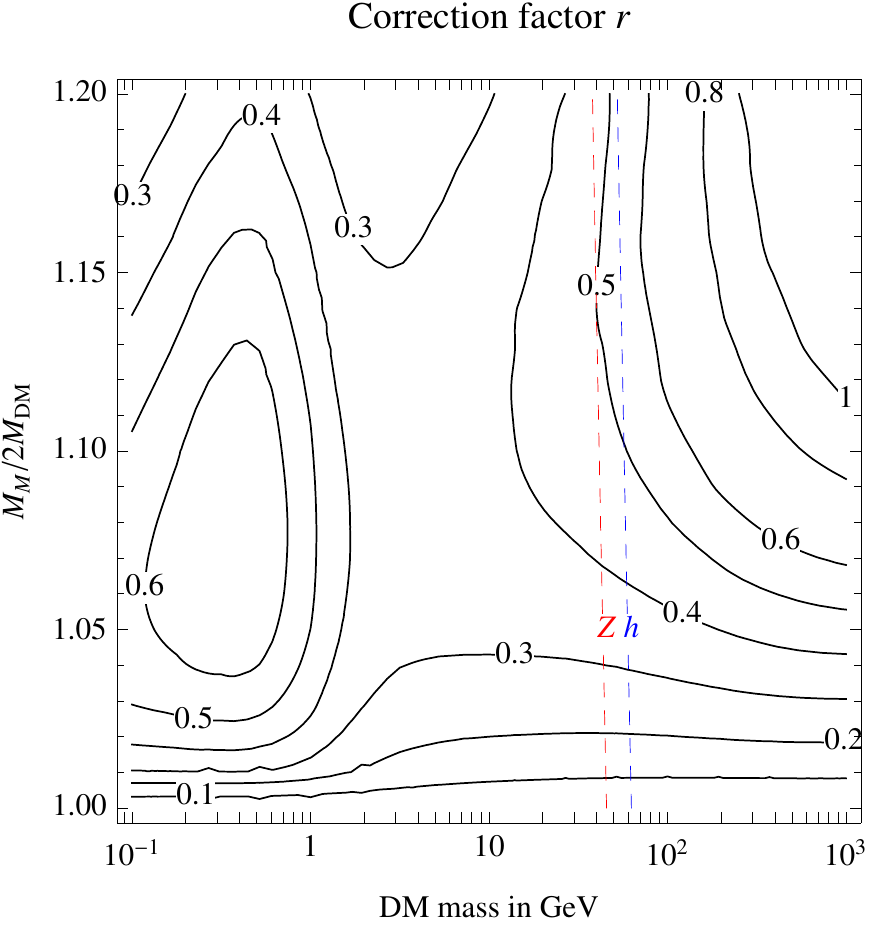}
\caption{\em {\bf DM freeze-out via decays} of a generic mediator particle $M$.
The dashed curves emphasise the special cases where the mediator is the $Z$ (red) or the Higgs boson (blue).
Left panel: value of $(g_{M}/g_{\rm DM}^2)\Gamma_{M\to {\rm DM}}/M_{\rm DM}$ such that
decays into DM particles reproduce the observed DM abundance.
Right panel: the order-one factor $r$ that gives the precise normalisation of the DM thermal abundance via decays,
as defined in eq.~(\ref{OmegaRes}).
\label{fig:DMres}} 
\end{figure}


%

\medskip

The total thermally averaged DM annihilation rate $\gamma_A$ can be decomposed as
the contribution of the on-shell resonant term 
plus the remaining off-shell contribution $\gamma_A^{\rm sub}$:
\beq \gamma_A =\gamma_A^{\rm on-shell} + \gamma_A^{\rm off-shell} .\eeq
Formally, $\gamma_A^{\rm on-shell} $ can be computed by approximating the Breit-Wigner as a Dirac $\delta$ function.
As expected, the scattering rate reduces to a much simpler object: the thermal average $\gamma_D$ of the
decay rate of the mediator,
\beq \gamma_A^{\rm on-shell}  =
\hbox{BR}_{\rm DM}(1-\hbox{BR}_{\rm DM}) \gamma_D.\eeq
The  term $\gamma_A^{\rm off-shell}$ can be computed using a subtracted propagator for the mediator particle $M$,
as described in~\cite{Leptog}.
However, in the context of section~\ref{sec3}, the off-shell contribution is model-dependent.

We focus on the model-independent on-shell term, which is described by the decay rate.
In this approximation one has the simple result
\beq \gamma_A \approx \hbox{BR}_{\rm DM}\gamma_D = n_M^{\rm eq} \frac{K_1(M_M/T)}{K_2(M_M/T)}\Gamma_{M\to{\rm DM}}\stackrel{T\ll M_{\rm DM}}
\simeq
g_M \Gamma_{M\to {\rm DM}} (\frac{M_M T}{2\pi})^{3/2} e^{-M_M/T}.
\eeq
We considered the relevant non-relativistic limit, and we
notice that $\gamma_D$ has a different dependence on $T$ than the standard annihilation rate
(proportional to $e^{-2M_{\rm DM}/T}$),
%
so the usual approximation in terms of the non-relativistic parameter $\sigma v$ is not appropriate.
Rather, the DM abundance is determined in terms of the width $\Gamma_{M\to DM}$,
such that the final DM number abundance is roughly given by $n_{\rm DM}/s\sim H/\Gamma_{M\to {\rm DM}}$
where $H$ is the Hubble constant at $T\sim M_{\rm DM}$.
By  solving the Boltzmann equation for the DM abundance keeping only the on-shell term
we find the precise result
\beq\label{OmegaRes}
  \frac{\Omega_{\rm DM} h^2}{0.1187}= r
\frac{g_{\rm DM}^2 10^{-12}\GeV}{g_M\Gamma_{M\to{\rm DM}}} \left(\frac{M_{\rm DM}}{\GeV}\right)^3
 e^{-2 z_f \Delta M/M_{\rm DM}} \, .
 \eeq
We have defined $\Delta M = M_M - 2 M_{\rm DM}$ 
and fixed $z_f \equiv 25$. Then
$r$ is an order-one  factor plotted in fig.~\ref{fig:DMres} (right panel),
obtained from the numerical solution of the Boltzmann equations.
As in the annihilation case, the relevant rate is averaged over the DM components and summed over the SM components. In the left panel of fig.~\ref{fig:DMres}, we show the invisible width of the mediator, in units of its mass, that corresponds to the correct DM abundance via decays.

\begin{figure}[t]
\centering
\includegraphics[width=0.65\textwidth]{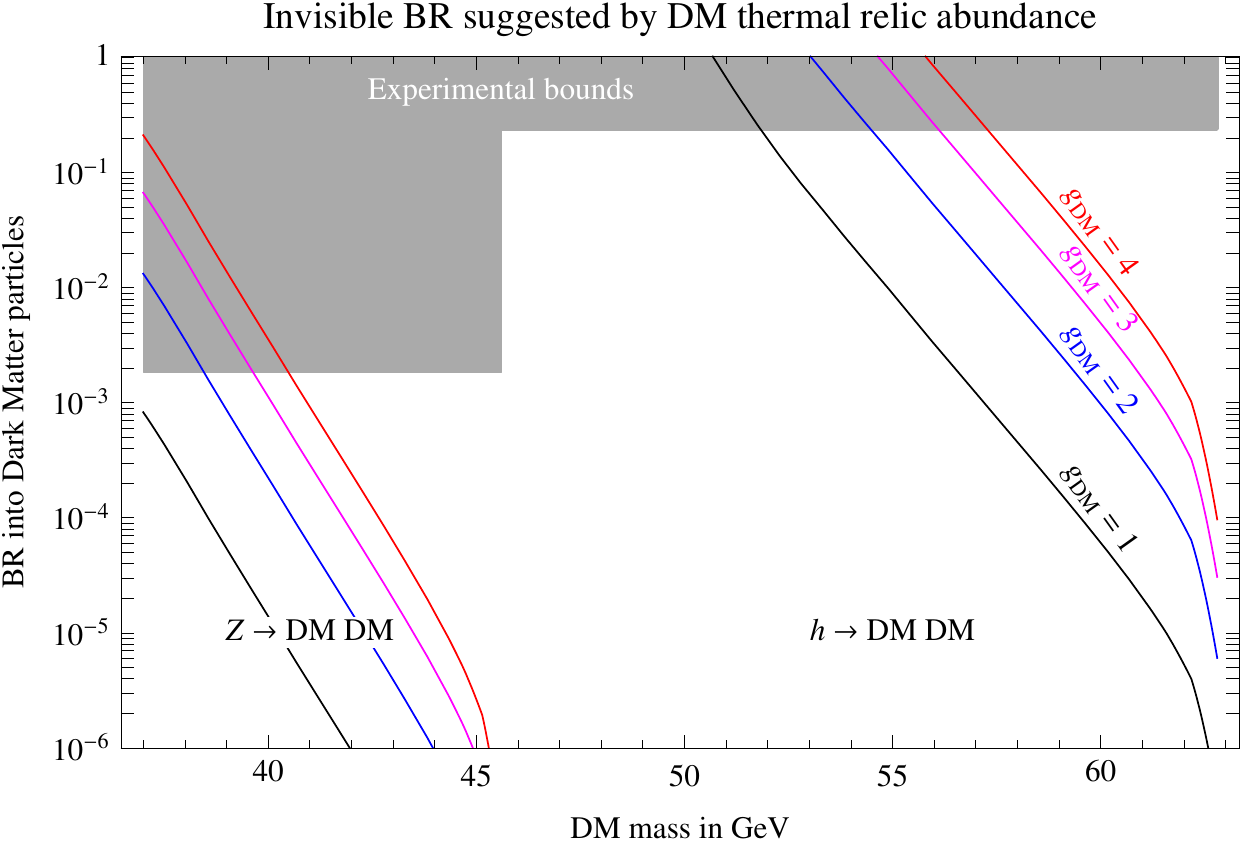}
\caption{\em Values of the invisible $Z$ and $h$ branching ratio into DM particles needed to
reproduce the DM abundance via thermal freeze-out of decays.
We consider DM with $g_{\rm DM}=1$ (black curve),
2 (blue), 3 (magenta), 4 (red)  degrees of freedom.
\label{fig:DMresZh}} 
\end{figure}

We can now apply our results to the case in which the mediator $M$ is either the $Z$ or the Higgs boson. For the various couplings considered in section~\ref{sec3}, the decay widths into DM particles are
\beq \Gamma_{Z\to{\rm DM}} = \frac{g_2^2 M_Z}{12\pi\cos^2\theta_{\rm W}} \sqrt{1 - \frac{4M_{\rm DM}^2}{M_Z^2}}\left\{
\begin{array}{l}
g_V^{\rm DM2}(1+2M_{\rm DM}^2/M_Z^2)\\
g_A^{\rm DM2} (1-4M_{\rm DM}^2/M_Z^2)\\
g_s^2(1/4-M_{\rm DM}^2/M_Z^2)
\end{array}\right.   \, ,
\eeq
\beq \Gamma_{h\to{\rm DM}} = \frac{M_h}{16\pi} \sqrt{1 - \frac{4M_{\rm DM}^2}{M_h^2}}\left\{
\begin{array}{l}
y_{\rm DM}^2(1-4M_{\rm DM}^2/M_h^2)\\
y_{\rm DM}^{P2}\\
\frac{1}{2}\lambda_{\rm DM}^2 (v/M_h)^2
\end{array}\right.\, .
\eeq
The values of the invisible branching ratios needed to reproduce the DM abundance are
shown in fig.~\ref{fig:DMresZh}.
This result holds as long as the on-shell contribution that we are considering dominates over
the neglected off-shell contribution, which occurs typically for $\Delta M  \circa{<} 0.2 M_{\rm DM}$. As shown in fig.~\ref{fig:DMresZh}, a broad range of experimentally unexplored $Z$ or Higgs invisible widths could account for DM via thermal freeze-out of decays. This result gives good motivations for improved measurements of the invisible width of the $Z$ boson ({\it e.g.} in GigaZ) and of the Higgs boson (in upcoming LHC data and in future Higgs factories).

\section{Summary}
\label{sec:conclusions}

The search for DM is one of the most exciting goals of the LHC. However, the path that DM hunters should follow is not obvious because of our ignorance about the nature of the DM and the lack of experimental evidence for new particles beyond the SM. In this situation, experiments at the LHC must pursue a diversified strategy of searches. In this paper we have considered some benchmark cases that offer prospects for DM discovery and that can guide experimental searches without full commitment to specific models.

\paragraph{\bf DM co-annihilating with a coloured partner.} As a first benchmark case, we have considered a situation in which the DM thermal relic abundance is determined by co-annihilation with a coloured particle. This possibility is very favourable for the LHC, because of the large QCD cross section for producing the DM partner. However, the near mass degeneracy with the DM particle makes this search at the LHC experimentally challenging, requiring either extra jets to tag the event and/or the identification of soft decay products. 
Co-annihilation with coloured partners can occur in many models with new particles at the weak scale, including supersymmetry. Interesting examples are the cases of near-degenerate neutralino-stop, neutralino-sbottom, or neutralino-gluino. However, even without making any model-dependent assumption, the case of DM co-annihilating with a coloured partner can be fully characterised by: the quantum numbers of the DM partner (spin and colour representation), its mass, and its mass difference with the DM particle. In terms of these parameters, one can determine the DM relic abundance and the signals at the LHC. An important result of our study is the calculation of the Sommerfeld corrections to the annihilation rates, taking into account the colour decomposition of the various initial states. As a byproduct, we obtain the correct expression of the Sommerfeld factor for gluino LSP annihilation. Our results for the DM relic abundance are summarised in figs.~\ref{fig:cross} and \ref{fig:comb}. We find that future LHC searches will be able to probe a large region of parameters that is still unexplored and that leads to a correct DM density. However, LHC cannot give a conclusive answer to the viability of DM co-annihilating with coloured partners. Indeed, a correct thermal relic density can be achieved even for DM masses as large as 5 TeV (for a fermion colour triplet partner) or 10 TeV (for a fermion colour octet partner). Future hadron colliders operating at 100~TeV energies are necessary to complete the exploration of these models.

\paragraph{\bf DM annihilating through a SM mediator.} In a  large class of models the DM particle is coupled to the SM sector only through the $Z$ or Higgs boson. 
In this case, the mass of the mediator is known, but we treat the couplings between the DM particle and the $Z$ or Higgs boson as free parameters. 
Although the thermal DM abundance is somehow model-dependent, we can compare the reach of the different experimental strategies. We find that, taking into account the LEP bound on the $Z$ invisible width and the LUX 2013 data, searches at the LHC for DM coupled to the $Z$ are not sufficiently competitive. The situation is more promising for DM coupled to the Higgs because, for $M_{\rm DM}<M_h/2$, searches for invisible Higgs decays are competitive with direct searches in underground experiments. Our findings are illustrated in figs.~\ref{fig:exclusionsZ} and \ref{fig:exclusionsHiggs}. In spite of the negative results for DM coupled to the $Z$, improvements at the LHC in all missing-energy channels are motivated, independently of the relic density prediction and the LUX 2013 results. Indeed, DM could have a non-thermal origin (evading the relic abundance constraint) or could have a clumped distribution in the galactic halo (weakening the constraints from direct DM searches).

\paragraph{\bf DM freeze-out via decays.} An especially interesting situation occurs when DM annihilates through a near-resonant mediator. We have shown that, for a DM mass slightly smaller than half the mediator mass, the relic abundance is determined in a model-independent way by the invisible width of the mediator. The cases of $Z$ or Higgs as mediators offer interesting applications to our results. As shown in fig.~\ref{fig:DMresZh}, $Z$ and Higgs invisible widths below their experimental limits are compatible with DM thermal abundance and with the LUX constraint. Thus, the search for invisible Higgs decays that can be performed at the LHC and at future facilities (and, possibly, improvements in the measurement of the $Z$ invisible width at GigaZ) offer very interesting ways to probe the nature of the DM.

\section*{Acknowledgments}

We thank E. del Nobile, P. Panci and A. Urbano for useful discussions.
ADS acknowledges partial support from the  European Union FP7  ITN INVISIBLES 
(Marie Curie Actions, PITN-GA-2011-289442).
This work was supported by the {\sc ESF} grant MTT8 and by the SF0690030s09 project
and by the European Programme PITN-GA-2009-237920 (UNILHC).

\appendix

\section{Colour tensor products}\label{app}
We give here some details about 
the decomposition of the $\bar\chi'_I \chi'_J\to g^ag^b$ scattering rate into two-body channels with given colour.
The total scattering amplitude is proportional to $A^{ab}_{IJ}=\{T^a, T^b\}_{IJ}$, for both fermion and scalar particles $\chi'$.
Here $(T^a)_{IJ}$ are the colour generators in the desired representation: $(T^a)_{ij}=\lambda^a_{ij}/2$ for triplets and $(T^a)_{bc}=-if^a_{bc}$ for octets.
 
 When $\chi'$ is a colour triplet, the rate has to be decomposed into the $3\otimes\bar 3 = 1\oplus 8$ channels.
 The total amplitude $A^{ab}|_i^k$ can be split into singlet and octet amplitudes according to
\beq
A^{ab}|_i^k=[{\bf 1}]^{ab}|_i^k+[{\bf 8}]^{ab}|_i^k \, ,
\label{amp33}
\eeq
\beq
[{\bf 1}]^{ab}|_i^k=\frac13 \delta_i^k A^{ab}|_m^m\, ,\qquad
[{\bf 8}]^{ab}|_i^k=A^{ab}|_i^k -\frac13 \delta_i^k A^{ab}|_m^m \, .
\eeq
When we take the modulus squared of the amplitude in \eq{amp33} and sum over colour indices, the interference terms vanish and we are left with the squares of the individual amplitudes, which are given by
\beq
 \frac 72 \sum_{abik}\left| [{\bf 1}]^{ab}|_i^k \right|^2 = \frac 75 \sum_{abik}\left| [{\bf 8}]^{ab}|_i^k \right|^2
= \sum_{abik}\left| A^{ab}|_i^k \right|^2
  \, .
\eeq
This result explains the factors in eq.~(\ref{eq:S3}).

\medskip

When $\chi'$ is a color octet, the rate has to be decomposed into the
$8\otimes 8=1_S\oplus 8_A\oplus 8_S\oplus 10_A\oplus \overline{10}_A\oplus 27_S$ channels.
The total amplitude can be written as $A_{ij}^{lk}$. The pair of fundamental and anti-fundamental indices
$(i,l)$, subject to a traceless condition ($A_{mj}^{mk}=0$), describe one octet; the pair of indices $(j,k)$ under the traceless condition ($A_{im}^{lm}=0$) describe the second octet; for the sake of readability, we drop the indices $a$ and $b$ of the final-state gluons. The total amplitude can be decomposed as
\beq
A_{ij}^{lk}=[{\bf 1_S}]_{ij}^{lk}+[{\bf 8_A}]_{ij}^{lk}+[{\bf 8_S}]_{ij}^{lk}+[{\bf 10_A}]_{ij}^{lk}+[{\bf \overline{10}_A}]_{ij}^{lk}+
[{\bf 27_S}]_{ij}^{lk} \, ,
\label{amp88}
\eeq
\begin{eqnarray}
{[{\bf 1}_{\bf S}]}_{ij}^{lk} &=& \frac18 A_{mn}^{nm} (\delta_i^k \delta_j^l - \frac13 \delta_i^l\delta_j^k)\\
{[{\bf 8}_{\bf A}]}_{ij}^{lk} &=& \frac16  \left[ \delta_j^l (A_{im}^{mk}-A_{mi}^{km}) -  \delta_i^k (A_{jm}^{ml}-A_{mj}^{lm})\right] \\
{[{\bf 8}_{\bf S}]}_{ij}^{lk} &=& \frac{1}{20} \left[\delta_i^k (A_{mj}^{lm}+A_{jm}^{ml})+\delta_j^k (A_{mi}^{lm}+A_{im}^{ml})+
\delta_i^l(A_{mj}^{km}+A_{jm}^{mk})+\delta_j^l(A_{mi}^{km}+A_{im}^{mk})\right]+\nonumber\\
&&+\frac14 (A_{ij}^{lk} - A_{ji}^{lk}-A_{ij}^{kl}+A_{ji}^{kl})+\frac{1}{60}(\delta_i^l \delta_j^k - 9\delta_i^k \delta_j^l)
A_{mn}^{nm}\\
{[{\bf 10}_{\bf A}]}_{ij}^{lk} &=&  \frac14 (A_{ij}^{lk}-A_{ji}^{lk}+A_{ij}^{kl}-A_{ji}^{kl})-\frac{1}{12}\left[ \delta_j^l (A_{im}^{mk}-A_{mi}^{km})+\right. \nonumber   \\ 
&&  \left. -
\delta_i^l(A_{jm}^{mk} - A_{mj}^{km})+\delta_j^k (A_{im}^{ml}-A_{mi}^{lm}) - \delta_i^k (A_{jm}^{ml}-A_{mj}^{lm})\right] \\
{[\overline{\bf 10}_{\bf A}]}_{ij}^{lk} &=& \frac14 (A_{ij}^{lk}+A_{ji}^{lk}-A_{ij}^{kl}-A_{ji}^{kl})-\frac{1}{12}\left[\delta_j^l (A_{im}^{mk}-A_{mi}^{km})+\right. \nonumber   \\
&&\left. +
\delta_i^l(A_{jm}^{mk} - A_{mj}^{km})-\delta_j^k (A_{im}^{ml}-A_{mi}^{lm}) - \delta_i^k (A_{jm}^{ml}-A_{mj}^{lm})\right] \\
{[{\bf 27}_{\bf S}]}_{ij}^{lk} &=&  \frac14 (A_{ij}^{lk}+A_{ji}^{lk}+A_{ij}^{kl}+A_{ji}^{kl})-
\frac{1}{20}[\delta_i^k (A_{mj}^{lm}+A_{jm}^{ml}) +
\delta_j^k (A_{mi}^{lm}+A_{im}^{ml}) +
\nonumber   \\
&&+
  \delta_i^l (A_{mj}^{km}+A_{jm}^{mk})+
 \delta_j^l(A_{mi}^{km}+A_{im}^{mk})] + \frac{1}{40}(\delta_i^l \delta_j^k + \delta_i^k \delta_j^l)A_{mn}^{nm}\, .
\end{eqnarray}
In the modulus squared of the amplitude in \eq{amp88} summed over colour indices, there are no interference terms and the only non-vanishing terms are
\beq
 6 \sum \left|[{\bf 1_S}]\right|^2=3 \sum \left|[{\bf 8_S}]\right|^2=2 \sum \left|[{\bf 27_S}]\right|^2 =\sum |A|^2 \, .
\eeq
This explains the factors in eq.~(\ref{eq:S8}).

\begin{multicols}{2}

\footnotesize


\end{multicols}


\begin{thebibliography}{nn}\bibitem{0403.004}
  A.~Birkedal, K.~Matchev and M.~Perelstein,
  Phys.\ Rev.\ D {70} (2004) 077701
  [\hhref{hep-ph/0403004}].
    


\bibitem{0503.117}
  J.~L.~Feng, S.~Su and F.~Takayama,
  Phys.\ Rev.\ Lett.\  {96} (2006) 151802
  [\hhref{hep-ph/0503117}].
  


\bibitem{0808.3384}
  M.~Beltran, D.~Hooper, E.~W.~Kolb and Z.~C.~Krusberg,
  Phys.\ Rev.\ D {80} (2009) 043509
  [arXiv:\hhref{0808.3384}].
    


\bibitem{0912.4511}
  Q.~-H.~Cao, C.~-R.~Chen, C.~S.~Li and H.~Zhang,
  JHEP {1108} (2011) 018
  [arXiv:\hhref{0912.4511}].
    


\bibitem{1002.4137}
  M.~Beltran, D.~Hooper, E.~W.~Kolb, Z.~A.~C.~Krusberg and T.~M.~P.~Tait,
  JHEP {1009} (2010) 037
  [arXiv:\hhref{1002.4137}].
    


\bibitem{1003.1912}
  P.~Agrawal, Z.~Chacko, C.~Kilic and R.~K.~Mishra,
  arXiv:\hhref{1003.1912}.
    


\bibitem{1005.1286}
  J.~Goodman, M.~Ibe, A.~Rajaraman, W.~Shepherd, T.~M.~P.~Tait and H.~-B.~Yu,
  Phys.\ Lett.\ B {695} (2011) 185
  [arXiv:\hhref{1005.1286}].
    


\bibitem{1005.3797}
  Y.~Bai, P.~J.~Fox and R.~Harnik,
  JHEP {1012} (2010) 048
  [arXiv:\hhref{1005.3797}].
    

  


\bibitem{1008.1591}
  J.~Fan, M.~Reece and L.~-T.~Wang,
  JCAP {1011} (2010) 042
  [arXiv:\hhref{1008.1591}].
    


\bibitem{1008.1783}
  J.~Goodman, M.~Ibe, A.~Rajaraman, W.~Shepherd, T.~M.~P.~Tait and H.~-B.~Yu,
  Phys.\ Rev.\ D {82} (2010) 116010
  [arXiv:\hhref{1008.1783}].
    


\bibitem{1009.0008}
  J.~Goodman, M.~Ibe, A.~Rajaraman, W.~Shepherd, T.~M.~P.~Tait and H.~-B.~Yu,
  Nucl.\ Phys.\ B {844} (2011) 55
  [arXiv:\hhref{1009.0008}].
    


\bibitem{1011.2310}
  K.~Cheung, P.~-Y.~Tseng and T.~-C.~Yuan,
  JCAP {1101} (2011) 004
  [arXiv:\hhref{1011.2310}].
    


\bibitem{1012.2022}
  J.~-M.~Zheng, Z.~-H.~Yu, J.~-W.~Shao, X.~-J.~Bi, Z.~Li and H.~-H.~Zhang,
  Nucl.\ Phys.\ B {854} (2012) 350
  [arXiv:\hhref{1012.2022}].
    


\bibitem{1103.0240}
  P.~J.~Fox, R.~Harnik, J.~Kopp and Y.~Tsai,
  Phys.\ Rev.\ D {84} (2011) 014028
  [arXiv:\hhref{1103.0240}].
    


\bibitem{1103.3289}
  J.~-F.~Fortin and T.~M.~P.~Tait,
  Phys.\ Rev.\ D {85} (2012) 063506
  [arXiv:\hhref{1103.3289}].
    


\bibitem{1104.1429}
  M.~R.~Buckley,
  Phys.\ Rev.\ D {84} (2011) 043510
  [arXiv:\hhref{1104.1429}].
    


\bibitem{1104.5329}
  K.~Cheung, P.~-Y.~Tseng and T.~-C.~Yuan,
  JCAP {1106} (2011) 023
  [arXiv:\hhref{1104.5329}].
    


\bibitem{1107.2048}
  J.~Wang, C.~S.~Li, D.~Y.~Shao and H.~Zhang,
  Phys.\ Rev.\ D {84} (2011) 075011
  [arXiv:\hhref{1107.2048}].
    


\bibitem{1107.2118}
  M.~T.~Frandsen, F.~Kahlhoefer, S.~Sarkar and K.~Schmidt-Hoberg,
  JHEP {1109} (2011) 128
  [arXiv:\hhref{1107.2118}].
    


\bibitem{1108.1196}
  A.~Rajaraman, W.~Shepherd, T.~M.~P.~Tait and A.~M.~Wijangco,
  Phys.\ Rev.\ D {84} (2011) 095013
  [arXiv:\hhref{1108.1196}].
    


\bibitem{1108.1800}
  G.~F.~Giudice, B.~Gripaios and R.~Mahbubani,
  Phys.\ Rev.\ D {85} (2012) 075019
  [arXiv:\hhref{1108.1800}].
    


\bibitem{1109.3516}
  P.~Agrawal, S.~Blanchet, Z.~Chacko and C.~Kilic,
  Phys.\ Rev.\ D {86} (2012) 055002
  [arXiv:\hhref{1109.3516}].
    


\bibitem{1109.4398}
  P.~J.~Fox, R.~Harnik, J.~Kopp and Y.~Tsai,
  Phys.\ Rev.\ D {85} (2012) 056011
  [arXiv:\hhref{1109.4398}].
    


\bibitem{1111.2359}
  J.~Goodman and W.~Shepherd,
  arXiv:\hhref{1111.2359}.
    


\bibitem{1111.2835}
  K.~N.~Abazajian, P.~Agrawal, Z.~Chacko and C.~Kilic,
  Phys.\ Rev.\ D {85} (2012) 123543
  [arXiv:\hhref{1111.2835}].
    


\bibitem{1112.5457}
  I.~M.~Shoemaker and L.~Vecchi,
  Phys.\ Rev.\ D {86} (2012) 015023
  [arXiv:\hhref{1112.5457}].
    


\bibitem{1201.0506}
  R.~Ding and Y.~Liao,
  JHEP {1204} (2012) 054
  [arXiv:\hhref{1201.0506}].
    


\bibitem{1201.3402}
  K.~Cheung, P.~-Y.~Tseng, Y.~-L.~S.~Tsai and T.~-C.~Yuan,
  JCAP {1205} (2012) 001
  [arXiv:\hhref{1201.3402}].
    


\bibitem{1202.2894}
  H.~An, X.~Ji and L.~-T.~Wang,
  JHEP {1207} (2012) 182
  [arXiv:\hhref{1202.2894}].
    


\bibitem{1203.1662}
  P.~J.~Fox, R.~Harnik, R.~Primulando and C.~-T.~Yu,
  Phys.\ Rev.\ D {86} (2012) 015010
  [arXiv:\hhref{1203.1662}].
    


\bibitem{1203.3542}
  A.~L.~Fitzpatrick, W.~Haxton, E.~Katz, N.~Lubbers and Y.~Xu,
  JCAP {1302} (2013) 004
  [arXiv:\hhref{1203.3542}].
    


\bibitem{1204.3839}
  M.~T.~Frandsen, F.~Kahlhoefer, A.~Preston, S.~Sarkar and K.~Schmidt-Hoberg,
  JHEP {1207} (2012) 123
  [arXiv:\hhref{1204.3839}].
    


\bibitem{1206.0640}
  V.~Barger, W.~-Y.~Keung, D.~Marfatia and P.~-Y.~Tseng,
  Phys.\ Lett.\ B {717} (2012) 219
  [arXiv:\hhref{1206.0640}].
  


\bibitem{1207.1431}
  M.~Garny, A.~Ibarra, M.~Pato and S.~Vogl,
  JCAP {1211} (2012) 017
  [arXiv:\hhref{1207.1431}].
    


\bibitem{1207.3971}
  M.~T.~Frandsen, U.~Haisch, F.~Kahlhoefer, P.~Mertsch and K.~Schmidt-Hoberg,
  JCAP {1210} (2012) 033
  [arXiv:\hhref{1207.3971}].
  


\bibitem{1208.4361}
  Y.~Bai and T.~M.~P.~Tait,
  Phys.\ Lett.\ B {723} (2013) 384
  [arXiv:\hhref{1208.4361}].
    


\bibitem{1208.4605}
  U.~Haisch, F.~Kahlhoefer and J.~Unwin,
  JHEP {1307} (2013) 125
  [arXiv:\hhref{1208.4605}].
    


\bibitem{1209.0231}
  N.~F.~Bell, J.~B.~Dent, A.~J.~Galea, T.~D.~Jacques, L.~M.~Krauss and T.~J.~Weiler,
  Phys.\ Rev.\ D {86} (2012) 096011
  [arXiv:\hhref{1209.0231}].
    


\bibitem{1210.0195}
  F.~P.~Huang, C.~S.~Li, J.~Wang and D.~Y.~Shao,
  Phys.\ Rev.\ D {87} (2013) 094018
  [arXiv:\hhref{1210.0195}].
    


\bibitem{1210.0525}
  R.~C.~Cotta, J.~L.~Hewett, M.~P.~Le and T.~G.~Rizzo,
  Phys.\ Rev.\ D {88} (2013) 116009
  [arXiv:\hhref{1210.0525}].
    


\bibitem{1212.2221}
  H.~An, R.~Huo and L.~-T.~Wang,
  Phys.\ Dark Univ.\  {2} (2013) 50
  [arXiv:\hhref{1212.2221}].
    


\bibitem{1212.3352}
  L.~M.~Carpenter, A.~Nelson, C.~Shimmin, T.~M.~P.~Tait and D.~Whiteson,
  arXiv:\hhref{1212.3352}.
    


\bibitem{1301.1486}
  A.~De Simone, A.~Monin, A.~Thamm and A.~Urbano,
  JCAP {1302} (2013) 039
  [arXiv:\hhref{1301.1486}].
    


\bibitem{1302.3619}
  N.~Zhou, D.~Berge and D.~Whiteson,
  arXiv:\hhref{1302.3619}.
    


\bibitem{1303.3348}
  H.~Dreiner, D.~Schmeier and J.~Tattersall,
  Europhys.\ Lett.\  {102} (2013) 51001
  [arXiv:\hhref{1303.3348}].
    


\bibitem{1303.6638}
  T.~Lin, E.~W.~Kolb and L.~-T.~Wang,
  Phys.\ Rev.\ D {88} (2013) 063510
  [arXiv:\hhref{1303.6638}].
    


\bibitem{1306.4107}
  H.~M.~Lee, M.~Park and V.~Sanz,
  arXiv:\hhref{1306.4107}.
    


\bibitem{1307.1129}
 B.~Bellazzini, M.~Cliche and P.~Tanedo,
  Phys.\ Rev.\ D {88} (2013) 083506
  [arXiv:\hhref{1307.1129}].
    


\bibitem{1307.2253}
  G.~Busoni, A.~De Simone, E.~Morgante and A.~Riotto,
  Physics Letters B 728C (2014)
  [arXiv:\hhref{1307.2253}].
    


\bibitem{1307.5740}
  Z.~-H.~Yu, Q.~-S.~Yan and P.~-F.~Yin,
  arXiv:\hhref{1307.5740}.
    


\bibitem{1307.6277}
  S.~Profumo, W.~Shepherd and T.~Tait,
  arXiv:\hhref{1307.6277}.
    


\bibitem{1307.8120}
  S.~Chang, R.~Edezhath, J.~Hutchinson and M.~Luty,
  arXiv:\hhref{1307.8120}.
    


\bibitem{1308.0592}
   H.~An, L.~-T.~Wang and H.~Zhang,
  arXiv:\hhref{1308.0592}.
    


\bibitem{1308.0612}
  Y.~Bai and J.~Berger,
  JHEP {1311} (2013) 171
  [arXiv:\hhref{1308.0612}].
    


\bibitem{1308.2679}
  A.~DiFranzo, K.~I.~Nagao, A.~Rajaraman and T.~M.~P.~Tait,
  JHEP {1311} (2013) 014
  [arXiv:\hhref{1308.2679}].
    
\bibitem{1312.5281} 
  A.~Alves, S.~Profumo and F.~S.~Queiroz,
  arXiv:\hhref{1312.5281}.

\bibitem{1308.6799}
  O.~Buchmueller, M.~J.~Dolan and C.~McCabe,
  arXiv:\hhref{1308.6799}.
    


\bibitem{1310.4491}
  U.~Haisch, F.~Kahlhoefer and E.~Re,
  JHEP {1312} (2013) 007
  [arXiv:\hhref{1310.4491}].
    


\bibitem{1310.6047}
  M.~A.~Fedderke, E.~W.~Kolb, T.~Lin and L.~-T.~Wang,
  JCAP01(2014)001
  [arXiv:\hhref{1310.6047}].
       


\bibitem{1311.5896}
  C.~Cheung and D.~Sanford,
  arXiv:\hhref{1311.5896}.
   


\bibitem{1311.6169}  
    N.~F.~Bell, Y.~Cai and A.~D.~Medina,
  arXiv:\hhref{1311.6169}.
    


\bibitem{1311.7131}
  U.~Haisch, A.~Hibbs and E.~Re,
  arXiv:\hhref{1311.7131}.
  

\bibitem{1312.0009}
  M.~B.~Krauss, S.~Morisi, W.~Porod and W.~Winter,
  arXiv:\hhref{1312.0009}.
 
\bibitem{1402.1275} 
  G.~Busoni, A.~De Simone, J.~Gramling, E.~Morgante and A.~Riotto,
  arXiv:\hhref{1402.1275}.


\bibitem{0011335}
  C.~P.~Burgess, M.~Pospelov and T.~ter Veldhuis,
  Nucl.\ Phys.\ B {619} (2001) 709
  [\hhref{hep-ph/0011335}].
    


\bibitem{0405097}
  H.~Davoudiasl, R.~Kitano, T.~Li and H.~Murayama,
  Phys.\ Lett.\ B {609} (2005) 117
  [\hhref{hep-ph/0405097}].
    


\bibitem{0605188}
 B.~Patt and F.~Wilczek,
  hep-ph/0605188.
    


\bibitem{0808.0255}
  S.~Andreas, T.~Hambye and M.~H.~G.~Tytgat,
  JCAP {0810} (2008) 034
  [arXiv:\hhref{0808.0255}].
    


\bibitem{0811.0393}
  V.~Barger, P.~Langacker, M.~McCaskey, M.~Ramsey-Musolf and G.~Shaughnessy,
  Phys.\ Rev.\ D {79} (2009) 015018
  [arXiv:\hhref{0811.0393}].
    


\bibitem{0909.0520}
  R.~N.~Lerner and J.~McDonald,
  Phys.\ Rev.\ D {80} (2009) 123507
  [arXiv:\hhref{0909.0520}].
    


\bibitem{0912.4722}
  X.~-G.~He, T.~Li, X.~-Q.~Li, J.~Tandean and H.~-C.~Tsai,
  Phys.\ Lett.\ B {688} (2010) 332
  [arXiv:\hhref{0912.4722}].
    


\bibitem{1005.5651}
  S.~Kanemura, S.~Matsumoto, T.~Nabeshima and N.~Okada,
  Phys.\ Rev.\ D {82} (2010) 055026
  [arXiv:\hhref{1005.5651}].
    


\bibitem{1008.1796}
  V.~Barger, Y.~Gao, M.~McCaskey and G.~Shaughnessy,
  Phys.\ Rev.\ D {82} (2010) 095011
  [arXiv:\hhref{1008.1796}].
    


\bibitem{1102.3024}
  A.~Biswas and D.~Majumdar,
  Pramana {80} (2013) 539
  [arXiv:\hhref{1102.3024}].
    


\bibitem{1106.3097}
 C.~Englert, T.~Plehn, D.~Zerwas and P.~M.~Zerwas,
  Phys.\ Lett.\ B {703} (2011) 298
  [arXiv:\hhref{1106.3097}].
    


\bibitem{1108.0671}
  Y.~Mambrini,
  Phys.\ Rev.\ D {84} (2011) 115017
  [arXiv:\hhref{1108.0671}].
    


\bibitem{1109.4872}
  M.~Pospelov and A.~Ritz,
  Phys.\ Rev.\ D {84} (2011) 113001
  [arXiv:\hhref{1109.4872}].
    


\bibitem{1110.4405}
  I.~Low, P.~Schwaller, G.~Shaughnessy and C.~E.~M.~Wagner,
  Phys.\ Rev.\ D {85} (2012) 015009
  [arXiv:\hhref{1110.4405}].
    


\bibitem{1111.4482}
  O.~Lebedev, H.~M.~Lee and Y.~Mambrini,
  Phys.\ Lett.\ B {707} (2012) 570
  [arXiv:\hhref{1111.4482}].
    


\bibitem{1112.3299}
  A.~Djouadi, O.~Lebedev, Y.~Mambrini and J.~Quevillon,
  Phys.\ Lett.\ B {709} (2012) 65
  [arXiv:\hhref{1112.3299}].
    


\bibitem{1201.4814}
 J.~F.~Kamenik and C.~Smith,
  Phys.\ Rev.\ D {85} (2012) 093017
  [arXiv:\hhref{1201.4814}].
    


\bibitem{1203.2064}
  L.~Lopez-Honorez, T.~Schwetz and J.~Zupan,
  Phys.\ Lett.\ B {716} (2012) 179
  [arXiv:\hhref{1203.2064}].
    


\bibitem{1205.3169}
  A.~Djouadi, A.~Falkowski, Y.~Mambrini and J.~Quevillon,
  Eur.\ Phys.\ J.\ C {73} (2013) 2455
  [arXiv:\hhref{1205.3169}].
    


\bibitem{1309.3561}
  A.~Greljo, J.~Julio, J.~F.~Kamenik, C.~Smith and J.~Zupan,
  JHEP {1311} (2013) 190
  [arXiv:\hhref{1309.3561}].
  

\bibitem{1311.1511}
  A.~A.~Petrov and W.~Shepherd,
   arXiv:\hhref{1311.1511}.

\bibitem{1312.2592} 
  L.~Carpenter, A.~DiFranzo, M.~Mulhearn, C.~Shimmin, S.~Tulin and D.~Whiteson,
  arXiv:\hhref{1312.2592}.

\bibitem{1401.0221} 
  G.~Arcadi, Y.~Mambrini, M.~H.~G.~Tytgat and B.~Zaldivar,
  JHEP {\bf 1403}, 134 (2014)
   [arXiv:\hhref{1401.0221}].  
   
\bibitem{1402.1173} 
  A.~Crivellin, F.~D'Eramo and M.~Procura,
\hhref{arXiv:1402.1173}.  
   
\bibitem{MDM}   
M.~Cirelli, N.~Fornengo and A.~Strumia,
  Nucl.\ Phys.\ B {753} (2006) 178
  [arXiv:\hhref{hep-ph/0512090}].
   
\bibitem{Sommerfeld} 
A.~Sommerfeld, Ann. Phys. {11} 257 (1931). 


\bibitem{9806361}
  H.~Baer, K.M.~Cheung and J.~F.~Gunion,
  Phys.\ Rev.\ D {59} (1999) 075002
  [\hhref{hep-ph/9806361}].
    


\bibitem{0307216}
  J.~Hisano, S.~Matsumoto and M.~M.~Nojiri,
  Phys.\ Rev.\ Lett.\  {92} (2004) 031303
  [\hhref{hep-ph/0307216}].
    


\bibitem{0412403}
  J.~Hisano, S.~.Matsumoto, M.~M.~Nojiri and O.~Saito,
  Phys.\ Rev.\ D {71} (2005) 063528
  [\hhref{hep-ph/0412403}].
  


\bibitem{Strumia:2008cf} 
  A.~Strumia,
  Nucl.\ Phys.\ B {809}, 308 (2009) {[arXiv:\hhref{0806.1630}]}.  
  


\bibitem{0810.0713}
  N.~Arkani-Hamed, D.~P.~Finkbeiner, T.~R.~Slatyer and N.~Weiner,
  Phys.\ Rev.\ D {79} (2009) 015014
  [arXiv:\hhref{0810.0713}].
    


\bibitem{1005.4678}
  J.~L.~Feng, M.~Kaplinghat and H.~-B.~Yu,
  Phys.\ Rev.\ D {82} (2010) 083525
  [arXiv:\hhref{1005.4678}].
   


\bibitem{Freitas}
  A.~Freitas,
  Phys.\ Lett.\ B {652} (2007) 280 {[arXiv:\hhref{0705.4027}]}.


\bibitem{Hryczuk}	  
 A.~Hryczuk,
  Phys.\ Lett.\ B {699} (2011) 271 [arXiv:\hhref{1102.4295}].  
 


\bibitem{VQCD1}
  W. Fischler, Nucl. Phys. B {129} (1977) 1157.


\bibitem{VQCD2}  
  Y. Schroder, Phys. Lett. B {447} (1999) 321 
         	  \href{http://arXiv.org/abs/hep- ph/9812205}{[hep- ph/9812205]}.


\bibitem{Pierini} 
  A.~Delgado, G.~F.~Giudice, G.~Isidori, M.~Pierini and A.~Strumia,
  Eur.\ Phys.\ J.\ C {73}, 2370 (2013), [arXiv:\hhref{1212.6847}].
  


\bibitem{madgraph} 
  J.~Alwall, M.~Herquet, F.~Maltoni, O.~Mattelaer and T.~Stelzer,
  JHEP {1106}, 128 (2011) [arXiv:\hhref{1106.0522}].
    
  
  
  


\bibitem{monojetATLAS}
\href{http://cds.cern.ch/record/1493486}{ATLAS-CONF-2012-147}.	


\bibitem{monojetCMS}	
	\href{http://cds.cern.ch/record/1525585}{CMS-PAS-EXO-12-048}

  


\bibitem{talk}
L. T. Wang, \href{https://indico.cern.ch/event/284800/session/0/contribution/5/material/slides/0.pdf}{talk at BSM Opportunities at 100 TeV}, CERN (2014).


\bibitem{Cirelli}
  E.~Del Nobile, M.~Cirelli and P.~Panci,
      	  \href{http://arXiv.org/abs/1307.5955}{arXiv:\hhref{1307.5955}}.
  
    

\bibitem{lux} 
  LUX Collaboration,
         	  \href{http://arXiv.org/abs/1310.8214}{arXiv:\hhref{1310.8214}}.
	


\bibitem{Haisch}
  U.~Haisch and F.~Kahlhoefer,
  JCAP {1304} (2013) 050
  [arXiv:\hhref{1302.4454}].


\bibitem{Leptog}
  G.~F.~Giudice, A.~Notari, M.~Raidal, A.~Riotto and A.~Strumia,
  Nucl.\ Phys.\ B {685} (2004) 89
 [\hhref{hep-ph/0310123}].


  
\end{thebibliography}
\end{document}